\documentclass[a4paper,11pt]{article}

\usepackage{float}
\usepackage{graphicx}
\usepackage[labelfont=bf,labelsep=period]{caption}
\usepackage{url}
\usepackage{xspace}
\usepackage[ruled]{algorithm}
\usepackage{refcount}
\usepackage{numbertabbing}
\usepackage{mathrsfs}
\usepackage{hyperref}

\usepackage[margin=2.5cm]{geometry}

\floatstyle{ruled}
\newfloat{algo}{htbp}{algo}
\floatname{algo}{Algorithm}
\usepackage{cite} 

\def \ifempty#1{\def\temp{#1} \ifx\temp\empty }

\newcommand{\str}[1]{\textsc{#1}}
\newcommand{\var}[1]{\textit{#1}}
\newcommand{\op}[1]{\textsl{#1}}

\newcommand{\false}{\textsc{false}\xspace}
\newcommand{\true}{\textsc{true}\xspace}

\usepackage{amsmath} 
\allowdisplaybreaks[2]          

\usepackage{amssymb} 

\usepackage{amsthm} 
\newtheoremstyle{plain-boldhead}
  {\topsep}
  {\topsep}
  {\itshape}
  {}
  {\bfseries}
  {.}
  { }
  {\thmname{#1}\thmnumber{ #2}\thmnote{ (\bfseries #3)}}
\newtheoremstyle{definition-boldhead}
  {\topsep}
  {\topsep}
  {\normalfont}
  {}
  {\bfseries}
  {.}
  { }
  {\thmname{#1}\thmnumber{ #2}\thmnote{ (\bfseries #3)}}
\theoremstyle{plain-boldhead}
\newtheorem{theorem}{Theorem}

\newtheorem{lemma}[theorem]{Lemma}

\theoremstyle{definition-boldhead}
\newtheorem{definition}{Definition}
\newtheorem{remark}{Remark}
\newtheorem{example}{Example}

\providecommand{\keywords}[1]
{
  \small	
  \textbf{Keywords.} #1
}

\begin{document}

\title{\bf Quorum Systems in Permissionless Networks}

\author{Christian Cachin\\
  University of Bern\\
  \url{cachin@inf.unibe.ch}
  \and Giuliano Losa\\
    Stellar Development Foundation\\
  \url{giuliano@stellar.org}
  \and Luca Zanolini\\
   University of Bern\\
  \url{luca.zanolini@unibe.ch}
}


\maketitle

\begin{abstract}
Fail-prone systems, and their quorum systems, are useful tools for the design of distributed algorithms.
However, fail-prone systems as studied so far require every process to know the full system membership in order to guarantee safety through globally intersecting quorums.
Thus, they are of little help in an open, permissionless setting, where such knowledge may not be available.
 We propose to generalize the theory of fail-prone systems to make it applicable to permissionless systems.
 We do so by enabling processes not only to make assumptions about failures, but also to make assumptions about the assumptions of other processes.
  Thus, by transitivity, processes that do not even know of any common process may nevertheless have intersecting quorums and solve, for example, reliable broadcast.
  Our model generalizes existing models such as the classic fail-prone system model [Malkhi and Reiter, 1998] and the asymmetric fail-prone system model [Cachin and Tackmann, OPODIS 2019]. 
  Moreover, it gives a characterization with standard formalism of the model used by the Stellar blockchain.
\end{abstract}

\keywords{Permissionless systems, fail-prone system, quorum system}

\section{Introduction}
\label{sec:introduction}

A common problem in distributed computing is to implement synchronization abstractions such as reliable broadcast, shared memory, or consensus, given some assumptions about the possible Byzantine failures that may occur in an execution.

A \emph{fail-prone system} $\mathcal{F}$~\cite{DBLP:journals/dc/MalkhiR98} is a set of sets of processes, called \emph{fail-prone sets}, where no fail-prone set is a subset of another.
A fail-prone system $\mathcal{F}$ denotes the assumption that the set of processes~$A$ that may suffer Byzantine failures is contained in one of the fail-prone sets.
For example, in a system of $n$ processes, it is common to assume that less than a third will fail,
i.e., the fail-prone sets are the sets of cardinality exactly $\lfloor (n-1)/3\rfloor$.


Fail-prone systems are useful because of their relationship to quorum systems~\cite{DBLP:journals/dc/MalkhiR98}.
A Byzantine quorum system $\mathcal{Q}$ for $\mathcal{F}$ is a collection of subsets of processes, called \emph{quorums}, such that for every two quorums $Q_1$ and $Q_2$ in $\mathcal{Q}$ and for every fail-prone set $F \in \mathcal{F}$ it holds that $Q_1$ and $Q_2$ have a common member outside $F$ (Consistency) and for every fail-prone set $F \in \mathcal{F}$, there exists a quorum disjoint from $F$ (Availability).

Many distributed algorithms (implementing, e.g., Byzantine reliable broadcast or consensus) are parameterized by a quorum system $\mathcal{Q}$ and their guarantees hold under the assumptions of a fail-prone system $\mathcal{F}$ if and only if Consistency and Availability of $\mathcal{Q}$ hold.
This allows the designers of a distributed system to make assumptions about failures, pick a
corresponding quorum system, and then choose among existing algorithms to solve the desired synchronization problem.

Traditionally, such fail-prone systems have been used in closed systems with assumptions of the form ``less than on third of the processes are faulty''.
This can work even in permissionless systems using Proof-of-Stake, e.g., assuming that less than one third of the stake-holders are faulty, or Proof-of-Work, e.g., assuming that the faulty set of processes holds less than one third of the mining power~\cite{DBLP:conf/wdag/PassS17}.
Proof-of-Stake and Proof-of-Work however have their disadvantages, e.g., long-range attacks in Proof-of-Stake, or excessive energy consumption in Proof-of-Work.
Both may also suffer from the ``rich getting richer'' problem, leading to very few entities eventually controlling the system.

Instead of Proof-of-Stake and Proof-of-Work, one can envision letting every participant make its own, subjective failure assumptions.
Distributed computing models based on this idea was first investigated by Damgård \emph{et al}.~\cite{DBLP:conf/asiacrypt/DamgardDFN07}, followed by Sheff et al.~\cite{DBLP:journals/corr/SheffRM14}, and more recently by Cachin and Tackmann~\cite{DBLP:conf/opodis/CachinT19} with the \emph{asymmetric-trust model}, Malkhi \emph{et al.}~\cite{DBLP:conf/ccs/MalkhiN019} with \emph{flexible quorums}, and Sheff et al.~\cite{DBLP:conf/opodis/SheffWRM20} with Heterogeneous Paxos.
On the practical side, Ripple~(\href{https://ripple.com}{https://ripple.com}) deployed a permissionless consensus protocol based on subjective quorums in 2012, in every process declares a \emph{Unique Node List} (UNL) of processes that it trusts. Stellar (\href{https://stellar.org}{https://stellar.org},~\cite{Mazieres2015TheSC}) later followed suite with a different permissioness model based on subjective trust.

One problem is that, in all the models cited above, even though participants are free to make their own failure assumptions or choose their own quorums, maintaining consistency requires compatible assumptions (in the sense that the resulting quorums will sufficiently intersect) and thus prior common knowledge or prior synchronization, which is not desirable in a permissionless system.


For example, Cachin and Tackmann~\cite{DBLP:conf/opodis/CachinT19} assume that for every two participants $p_i$ and $p_j$, for every two quorums $Q_i$ of $p_i$ and $Q_j$ of $p_j$, if $F$ is a set that can fail according to the assumptions of both $p_i$ and $p_j$, then $\left(Q_i\cap Q_j\right) \setminus F\neq \emptyset$; this is called the \emph{consistency} property. Together with an \emph{availability} property, this defines an \emph{asymmetric quorum system}.
Consistency ensures that if both $p_i$ and $p_j$ make correct assumptions, then they can avoid disagreeing in, e.g., a reliable broadcast protocol. Achieving liveness additionally requires the availability property, i.e., that a group of participants, called a \emph{guild}, whose assumptions are correct and which do not fail, satisfy the consistency property (pairwise) and additionally that every member of the guild has a quorum in the guild.

Maintaining safety in such a model requires a strong form of a-priori coordination.
Indeed, two participants cannot be prevented from disagreeing unless every two of their quorums have at least one non-faulty participant in common.
Thus participants must coordinate beforehand in order to pick sufficiently overlapping survivor sets, and corresponding quorums.
In the model of Ripple, for instance, every two processes must have UNLs that overlap by some sufficient fraction~\cite{DBLP:conf/opodis/Amores-SesarCM20}.

Global assumptions implying the intersection of survivor sets and quorums, as in the two preceding examples, are problematic in a permissionless setting because they postulate some form of pre-agreement or common knowledge, which might be hard to achieve in practice.

Interestingly, the Stellar network~(\href{https://stellar.org}{https://stellar.org},~\cite{Mazieres2015TheSC}), a deployed blockchain system based on quorums, is able to maintain safety and liveness without requiring that participants choose intersecting quorums.
Instead, participants choose \emph{quorum slices} that need not intersect, and the quorums of a participant are defined in terms of the slices of other participants.
Consensus can then be solved within an \emph{intact set} $\mathcal{I}$, a set of correct processes such that every two processes $p_i$ and $p_j$ in $\mathcal{I}$ have all their own quorums that intersect in at least a member of $\mathcal{I}$ and such that $\mathcal{I}$ is itself a quorum for every of its members.

We observe that quorum slices can be interpreted as a new kind of failure assumptions:
a participant assumes that at least one of its quorum slices is made exclusively of participants that do not fail and make correct assumptions.
In other words, a participant's assumption are not only about failures, but also about whether other participants make correct assumptions.
In practice, the Stellar model makes it easier for participants to achieve quorum intersection by relying on the failure assumptions of other participants that might have more knowledge about the system than they have.

The main contribution of this paper is to show that this new kind of failure assumptions yield a generalization of the theory of fail-prone systems (i.e., classic fail-prone systems are a special case) which allows to obtain intersecting quorums even when participants do not know any common third party.

Moreover, based on this, we introduce the notion of \emph{permissionless fail-prone system} from which it is possible to derive a \emph{permissionless quorum system.}

This paper is structured as follows.
 In Section~~\ref{sec:model} we formally define the assumptions of a process; each process assumes that one of its slices $S$ will not fail and, \emph{additionally}, that the assumptions of every process in $S$ will be satisfied too.
Then, we introduce the notion of permissionless fail-prone system.
This extends and generalizes the \emph{asymmetric fail-prone system}~\cite{DBLP:conf/asiacrypt/DamgardDFN07, DBLP:conf/opodis/CachinT19}.
Crucially, we note that the new meaning of the assumptions of the processes allows processes to transitively rely on the assumptions of other processes.
However, this in turn enables malicious processes to lie about their assumptions.

In Section~\ref{sec:permissionless-quorum-system}, we propose a computation model, based on the notion of \emph{view}, that takes this phenomenon into account.
Moreover, always in Section~\ref{sec:permissionless-quorum-system}, we derive from the permissionless fail-prone system the notion of \emph{permissionless Byzantine quorum system}.

In Section~\ref{sec:leagues} we introduce the notion of \emph{league}, which is a set of processes $L$ that enjoys Consistency (quorums intersection) and Availability (existence of a quorum in $L$ consisting of correct processes) among the correct members of $L$ even when faulty processes lie about their configuration.

We compare our permissionless model with the classic model based on fail-prone systems~\cite{DBLP:journals/dc/MalkhiR98}, with the asymmetric model\cite{DBLP:conf/asiacrypt/DamgardDFN07, DBLP:conf/opodis/CachinT19}, with the federated Byzantine agreement system~\cite{Mazieres2015TheSC} and with the personal Byzantine quorum system model~\cite{DBLP:conf/wdag/LosaGM19} in Section~\ref{sec:comparison}. 
Interestingly we show that classic fail-prone systems can be understood as a special case of our model.

{In Section~\ref{sec:shared-memory} we present a first application of permissionless quorum systems through the emulation of shared memory, represented by a register. 
In particular we show how to implement a single-writer multi-reader register with permissionless quorum systems.}

{In Section~\ref{sec:reliable-broadcast} we show how a traditional synchronization protocol, i.e., Bracha broadcast~\cite{DBLP:journals/iandc/Bracha87}, can be adapted to work in our model, thereby offering a new toolbox for the design of permissionless distributed systems.}

Related work and conclusions are presented in Section~\ref{sec:related-work} and Section~\ref{sec:conclusions}, respectively. Finally, our model also leads to a characterization of the Stellar model with standard formalism~\cite{DBLP:journals/dc/MalkhiR98, DBLP:conf/asiacrypt/DamgardDFN07, DBLP:conf/opodis/CachinT19}.

\section{Model}
\label{sec:model}

We consider an unbounded set of \emph{processes} $\Pi=\left\{p_1,p_2,...\right\}$ that communicate asynchronously with each other by sending messages. 
We assume that processes do not necessarily know which other processes are in the system (i.e., each process only knows a subset of $\Pi$).

Processes are assigned a protocol to follow. Protocols are presented in a
modular way using the event-based notation of Cachin \emph{et al.}~\cite{DBLP:books/daglib/0025983}.
A process that follows its algorithm during an execution is called \emph{correct}.
Initially, all processes are correct, but a process may later fail, in which case it is called \emph{faulty}.
We assume Byzantine failures, where a process that fails thereafter behaves arbitrarily.

We assume that point-to-point communication between any two processes (that know each other) is available, as well as a best-effort gossip primitive that will reach all processes.
In a protocol, this primitive is accessed through the events ``sending a message through gossip'' and ``receiving a gossiped message.'' 
We assume that messages from correct processes to correct process are eventually received and cannot be forged.
The system itself is asynchronous, i.e., the delivery of messages among processes may be delayed arbitrarily and the processes have no synchronized clocks.

Processes make failure assumptions about other processes.
However, since a process does not know exactly who is part of the system, it cannot make failure assumptions about the whole system.
Instead, each process $p_i$ makes assumptions about a set $P_i \subseteq \Pi$, called $p_i$'s \emph{trusted set}, using a \emph{fail-prone system}~$\mathcal{F}_i$ over $P_i$ (Section \ref{def:fail-prone-system}).
Here, $\mathcal{F}_i$ is a collection of subsets of $P_i$ and $p_i$ believes that up to any set $F \in \mathcal{F}_i$ may jointly fail.
We say that $P_i$ and $\mathcal{F}_i$ constitute $p_i$'s \emph{assumptions} and they remain fixed during an execution.

Note that assuming that the assumptions of the processes are fixed is a simplification.
In practice, the system may experience churn, i.e., processes frequently entering and departing the system.
Processes that remain in the system can adjust their assumptions in response to churn: for example, if a process stops responding (maybe because it has left the system), then other processes can remove it from their assumptions.
Conversely, if a new process joins the system, existing processes may adjust their assumptions to include that process.
However, analyzing the system under churn is out of the scope of this paper.

\begin{definition}[Fail-prone system]
  \label{def:fail-prone-system}
  A \emph{fail-prone system} $\mathcal{F}$ over $P\subseteq \Pi$ is a set of subsets of $\Pi$ called \emph{fail-prone sets}, none of which contain the other (i.e., if $F\in\mathcal{F}$ and $F'\subset F$, then $F'\not\in\mathcal{F}$).
\end{definition}

A \emph{permissionless fail-prone system} (abbreviated PFPS) describes the assumptions of all the processes:

\begin{definition}[Permissionless fail-prone system]
  A \emph{permissionless fail-prone system} is an array $
    \mathbb{F}=\left[\left(P_1,\mathcal{F}_1\right),\left(P_2,\mathcal{F}_2\right),\ldots\right] $
    that associates each process $p_i$ to a trusted set $P_i\subseteq \Pi$ and a fail-prone system $\mathcal{F}_i$ over $P_i$. We refer to $(P_i, \mathcal{F}_i)$ as the \emph{configuration of process $p_i$}.
\end{definition}
We now consider a fixed PFPS $\mathbb{F}$.

\begin{definition}[Tolerated execution and tolerated set]
  \label{def:assms-satisfied}
We say that \emph{the assumptions of a process $p_i$ are satisfied} in an execution if the set $A$ of processes that actually fail is such that there exists a fail-prone set $F\in \mathcal{F}_i$ and:
  \begin{enumerate}
    \item $A\cap P_i\subseteq F$; and
    \item the assumptions of every member of $P_i\setminus F$ are satisfied.
  \end{enumerate}

If $p_i \in \Pi$ has its assumptions satisfied in an execution $e$, we say that $p_i$ \emph{tolerates the execution} $e$.

Finally, a set of processes $L$ \emph{tolerates a set} of processes $A$ if and only if every process $p_i \in L \setminus A$ tolerates an execution $e$ with set of faulty processes $A$.    
  
\end{definition}

\begin{example}
\label{ex:tolerated}
  Consider a set of processes $\Pi= \{p_1,p_2,p_3,p_4\}$ and a permissionless fail-prone system $\mathbb{F}=[(\Pi,\mathcal{F}_1),(\Pi,\mathcal{F}_2),(\Pi,\mathcal{F}_3),(\Pi,\mathcal{F}_4)]$ with $\mathcal{F}_1=\{\{p_3,p_4\}\}$, $\mathcal{F}_2=\{\{p_1,p_4\}\}$, $\mathcal{F}_3=\{\{p_1,p_4\}\}$, and $\mathcal{F}_4=\{\{p_1,p_2\}\}$.
  Then, $\Pi$ tolerates the sets $\emptyset$, $\{p_1\}, \{p_4\}$ and $\{p_1, p_4\}$.  
  To see this, let us assume an execution $e$ with set of faulty processes $A = \{p_1,p_4\}.$ Then, for every $p_i \in \Pi \setminus A$, there exists a fail-prone set $F \in \mathcal{F}_i$ such that $A \cap \Pi \subseteq F$. In particular, $\Pi \setminus A = \{p_2,p_3\}$ and  $\{p_1,p_4\} \in \mathcal{F}_2$ and $\{p_1,p_4\} \in \mathcal{F}_3$.
  The same reasoning can be applied for the other sets.
\end{example}

Note that here we depart significantly from the traditional notion of fail-prone systems~\cite{DBLP:journals/dc/MalkhiR98,DBLP:conf/opodis/CachinT19}: in a PFPS, a process not only makes assumptions about failures, but also makes assumptions about the assumptions of other processes.

Next we define \emph{survivor sets} analogously to Junqueira and Mar\-zul\-lo~\cite{DBLP:conf/icdcs/JunqueiraM03}.
In the traditional literature, a survivor set of $p_i$ is the complement, within $\Pi$, of some fail-prone set. However, defining them as the complement of fail-prone sets within $P_i$ does not work because of Item~2 in Definition~\ref{def:assms-satisfied}. To obtain this definition, we first define a \emph{slice}.

\begin{definition}[Slice]
\label{def:slice}
A set $\overline{F} \subseteq \Pi$ is a \emph{slice} of $p_i$ if and only if $p_i$ has a fail-prone set $F \in \mathcal{F}_i$ such that $\overline{F}=P_i\setminus F$.
\end{definition}

For any $S \subseteq \Pi$ we often say \emph{$p_i$ has a slice in~$S$}
when a slice of $p_i$ is contained in $S$ or when $S$
contains a superset of a slice of $p_i$.

\begin{definition}[Survivor-set system]
  \label{def:survset}
  A \emph{survivor-set system} $\mathcal{S}_i$ of $p_i$ is the minimal set of
  subsets $S$ of $\Pi$ such that:
  \begin{enumerate}
  \item $p_i$ has a slice in $S$; and
  \item every member of $S$ has a slice in $S$.
  \end{enumerate}
  Each $S \in \mathcal{S}_i$ is called a \emph{survivor set} of $p_i$.
\end{definition}

\begin{example}
\label{ex:slice-and-survivor}
Continuing from Example~\ref{ex:tolerated}, process $p_1$ has only one slice consisting of $\{p_1,p_2\}$, processes $p_2$ and $p_3$ have the set $\{p_2,p_3\}$ as slice, and process $p_4$ has the set $\{p_3,p_4\}$ as slice.
Moreover, the survivor-set systems are $\mathcal{S}_1 = \{\{p_1,p_2,p_3\}, \{p_1,p_2,p_3,p_4\}\}$ for process $p_1$, $\mathcal{S}_2 = \{\{p_2,p_3\}, \{p_1,p_2,p_3,p_4\}\}$ for process $p_2$, $\mathcal{S}_3 = \{\{p_2,p_3\}, \{p_1,p_2,p_3,p_4\}\}$ for process $p3$, and \sloppy{$\mathcal{S}_4 = \{\{p_2,p_3,p_4\}, \{p_1,p_2,p_3,p_4\}\}$ for process $p_4$.} This follows from Definition~\ref{def:survset}: given a survivor set $S \in \mathcal{S}_i$ for $p_i$, process $p_i$ must have a slice in $S$ and every member of $S$ must have a slice in $S$. So, for example, given the survivor set $\{p_1, p_2, p_3\}$ in the survivor set system $\mathcal{S}_1$ for $p_1$, process $p_1$ has a slice in $\{p_1, p_2, p_3\}$, i.e., $\{p_1, p_2\}$, and every process $p_i \in \{p_1, p_2, p_3\}$ has a slice in $ \{p_1, p_2, p_3\}$, i.e., $\{p_2,p_3\}$.

\end{example}

\begin{lemma}
  \label{lem:survivor-sets}
The assumptions of a process $p_i \in \Pi$ are satisfied in an execution $e$ with set of faulty processes $A$ if and only if there exists a survivor set $S\in \mathcal{S}_i$ of $p_i$ such that $S$ does not fail.  
\end{lemma}

\begin{proof} 
Let $p_i$ be a process such that, given an execution $e$ with set of faulty processes $A$, the assumptions of $p_i$ are satisfied in $e$.
This implies that, by Definition~\ref{def:assms-satisfied}, there exists a set of processes such that each of these processes has its assumptions satisfied. 
Moreover, by Definition~\ref{def:slice}, each of these processes has a slice $\overline{F}_j$ such that $\overline{F}_j \cap A = \emptyset$.
This leads to have a set $S$ obtained as union of all of these slices such that $S \cap A =\emptyset$ and such that $S$ is minimal with respect to this union, in the sense that is the minimal set of processes such that every process in $S$ has its assumptions satisfied. The set $S$ is a survivor set of $p_i$.

Conversely, we show that given a survivor set $S$ of $p_i$, given a process $p_i \in S$ and given an execution $e$ with set of faulty processes $A$, if $S \cap A =\emptyset$, then the assumptions of $p_i$ are satisfied in $e$.
Observe that, from the assumptions, we have that every process in $S$ has a slice $\overline{F}$ in $S$ such that $\overline{F} \cap A = \emptyset$. This means that for every process $p_i$ in $S$, there exists a fail-prone set $F \in \mathcal{F}_i$ such that $P_i \cap A \subseteq F$.
This implies that every process in $S$ has its assumptions satisfied and, in particular, that $p_i \in S$ has its assumptions satisfied in $e$.
\end{proof}

\section{Permissionless Quorum Systems}
\label{sec:permissionless-quorum-system}

A classic (or symmetric) fail-prone system~\cite{DBLP:journals/dc/MalkhiR98} determines a canonical quorum system known to all processes through the $Q^3$-condition.
Specifically, given a fail-prone system $\mathcal{F}$, the $Q^3$-condition requires that no three fail-prone sets of $\mathcal{F}$ cover the complete set of processes and this condition holds if and only if there exists a quorum system for $\mathcal{F}$~\cite{DBLP:journals/joc/HirtM00,DBLP:journals/dc/MalkhiR98}.
Such a quorum system could be, for example, the complement of every fail-prone set of $\mathcal{F}$, which we call the \emph{canonical} quorum system.
Traditional algorithms such as read-write register emulations~\cite{DBLP:journals/dc/MalkhiR98}, Byzantine reliable broadcasts~\cite{DBLP:journals/dc/SrikanthT87, DBLP:journals/iandc/Bracha87} or the PBFT algorithm~\cite{DBLP:journals/tocs/CastroL02} make use of quorums.

In the model of asymmetric trust~\cite{DBLP:conf/asiacrypt/DamgardDFN07} the assumptions of the processes may differ, and asymmetric quorum systems~\cite{DBLP:conf/opodis/CachinT19, DBLP:conf/esorics/CachinZ21} allow to implement the above-mentioned algorithms in a more flexible way. However, they still require a system that is known to every process.

In a permissionless system, processes do not know the membership and have different, partial, and potentially changing views of its composition.

Given a PFPS, we would therefore like to obtain a quorum system to implement algorithms for register emulation, broadcast, consensus and more, while allowing the processes to have different assumptions in an open network.

We are therefore interested in defining a notion of quorums for open systems where:
\begin{enumerate}
  \item each process has its own quorum system; and
  \item the quorums of a process $p_i$ depend on the assumptions of other processes, which $p_i$ learns by communicating with them.
\end{enumerate}

In other words, we consider scenarios in which each process $p_i$ communicates with other processes, continuously discovers new processes and learns their assumptions.
During this execution, $p_i$ determines its current set of quorums as a function of what it has learned so far.
Importantly, this means that the quorums of a process evolve as the process learns new assumptions, and that faulty processes can influence $p_i$'s quorums by lying about their assumption.

We now formalize this model using the notions of a \emph{view} and a \emph{quorum function}.

\begin{definition}[View]
  \label{def:view}
A view $\mathbb{V} = [\mathcal{V}_1,\mathcal{V}_2,\ldots]$ is an array with one entry $\mathbb{V}[j]=\mathcal{V}_j$ for each process $p_j$ such that:
\begin{enumerate}
  \item either $\mathcal{V}_j$ is the special value $\bot$; or
  \item $\mathcal{V}_j = (P_j, \mathcal{F}_j)$ consists of a set of processes $P_j$ and a fail-prone system $\mathcal{F}_j$ over $P_j$.
\end{enumerate}
\end{definition}

Observe that every process $p_i$ has its \emph{local} view $\mathbb{V}$, whose non-$\bot$ entries represent the assumptions that $p_i$ has learned at some point in an execution.
Every other process $p_j$ such that $\mathbb{V}[j]=\bot$ is a process that $p_i$ has not heard from.
We denote with $\Upsilon$ the set of all the possible views.

We assume that, for every process $p_j$, a process $p_i$'s view contains the assumption that $p_i$ has most recently received from $p_j$.
Finally, note that $\mathbb{F}$ is a view in which no process is mapped to~$\bot$.
In particular, $\mathbb{F}$ represents the global view if the system could be entirely observed.
Since processes cannot observe the complete system, they normally only have partial knowledge of $\mathbb{F}$.
Moreover, this knowledge evolves over time.

\begin{definition}[Domain of a view]
  For a view $\mathbb{V}$, the set of processes $p_i$ such that $\mathbb{V}[i]\neq \bot$ is the \emph{domain} of $\mathbb{V}$.
\end{definition}

Next, we assume that every process determines its quorums according to its view using a function $\mathscr{Q}$ called a \emph{quorum function}.
We assume that all correct processes use the same $\mathscr{Q}$ and that they do not change it during an execution.
We then have the following definition.

\begin{definition}[Quorum function]
  \label{def:quorum-function}
The \emph{quorum function} $\mathscr{Q}: \Pi \times \Upsilon \rightarrow 2^{\Pi}$ maps a process $p_i$ and a view $\mathbb{V}$ to a set of sets of processes such that $Q\in \mathscr{Q}(p_i,\mathbb{V})$ if and only if:
  \begin{enumerate}
  \item a slice of $p_i$ is contained in $Q$; and
  \item for every process $p_j\in Q$ with $\mathbb{V}[j] \neq \bot$
    and $\mathbb{V}[j] = \left(P_j,\mathcal{F}_j\right)$, there exists
    $F\in\mathcal{F}_j$ such that $P_j\setminus F\subseteq Q$.
  \end{enumerate}
  Every element of $\mathscr{Q}(p_i,\mathbb{V})$ is called a \emph{quorum for~$p_i$} (in $\mathbb{V}$).
\end{definition}

Notice that in the first condition, the quorum $Q$ may itself be a slice of
$p_i$.  Moreover, $Q$ is a quorum for every one of its members and it is
defined by slices of every $p_i \in Q$. As shown in the following lemma, a
quorum for $p_i$ in view $\mathbb{V}$ for $p_i$ is a survivor set of $p_i$.

\begin{lemma}
\label{lem:S-is-quorum}
For every view $\mathbb{V}$ for $p_i \in \Pi$, every quorum $Q_i \in \mathscr{Q}(p_i, \mathbb{V})$ is a survivor set of~$p_i$.
Moreover, given $S$ a survivor set of $p_i$, there exists a view $\mathbb{V}$ for $p_i$ such that $S \in \mathscr{Q}(p_i, \mathbb{V})$.
\end{lemma}

\begin{proof}
Let $Q_i \in \mathscr{Q}(p_i, \mathbb{V})$ be a quorum for $p_i$ with $\mathbb{V}$ a view for $p_i$.
By Definition~\ref{def:quorum-function}, all processes in $Q$ including $p_i$ have a slice in $Q$.
From Definition~\ref{def:survset}, this implies that $Q$ is a survivor set of~$p_i$.

Moreover, given a survivor set $S$ of $p_i$, the set $S$ consists of slices of every member of $S$. This means that there exists a view $\mathbb{V}$ for $p_i$ in which $S$ satisfies Definition~\ref{def:quorum-function} and it is a quorum for~$p_i$.
This is the view $\mathbb{V}$ defined as follows:
  \begin{enumerate}
    \item for every $p_j \in S$, $\mathbb{V}'[j]=\mathbb{F}[j]$, and 
    \item for every $p_j\not\in S$, $\mathbb{V}'[j]=(\emptyset, \{\emptyset\})$.
  \end{enumerate}
\end{proof}

\begin{example}
\label{ex:quorums}
Let us consider Example~\ref{ex:tolerated} with survivor-set systems as shown in Example~\ref{ex:slice-and-survivor}. 
Since all the processes already know all the configurations of every other process, we have that $\mathcal{S}_i = \mathscr{Q}(p_i, \mathbb{F})$, with $\mathbb{F}$ the permissionless fail-prone system.
\end{example}

Combining the quorum sets of all processes, we now obtain a \emph{permissionless quorum system} for $\mathbb{F}$.

\begin{definition}[Permissionless quorum system]
  \label{def:perm-quorum-system}
\sloppy{  A \emph{permissionless quorum system} for $\Pi$ and $\mathbb{F}$ is an array of collections of sets $\mathbb{Q}_{\text{perm}} = [\mathscr{Q}(p_1, \mathbb{F}), \mathscr{Q}(p_2, \mathbb{F}), \ldots]$,} where $\mathscr{Q}(p_i, \mathbb{F})$ is called the \emph{quorum system} for $p_i$ and is determined by the quorum function $\mathscr{Q}$.
\end{definition}

Observe that our notion of a quorum system differs from that in the existing literature~\cite{DBLP:journals/dc/MalkhiR98, DBLP:journals/siamcomp/MalkhiRW00, DBLP:conf/opodis/CachinT19}.
In particular, standard Byzantine quorum systems are defined through a pair-wise intersection among quorums.
This is possible in scenarios where the full system membership is known to every process.
However, in permissionless settings, this requirement cannot as clearly be achieved globally.


\begin{definition}[Current quorum system]
  Let $\mathbb{V}$ be the view representing the assumptions that a process $p_i$ has learned so far.
  Then the \emph{current quorum system of $p_i$} is the set $\mathscr{Q}(p_i, \mathbb{V})$.
  Moreover, a set of processes $Q$ is a \emph{current quorum} of $p_i$ if and only if $Q\in \mathscr{Q}(p_i, \mathbb{V})$; we also say that $p_i$ \emph{has a quorum}~$Q$.
\end{definition}

Note that, in this model, each process has its own set of quorums and the set of quorums of a process changes throughout an execution as the process learns the assumptions of more processes.
Importantly, note that faulty processes may lie about their configuration and influence the quorums of correct processes.
In an execution $e$ with faulty set $A$, a correct process $p_i$ might have a view in which the assumptions of processes in $A$ are arbitrary because processes in $A$ lied about their assumptions.
However, processes outside $A$ do not lie about their assumptions.
We capture this with the following definition.

\begin{definition}[T-resilient view]
\label{def:S-resilient}
  Given a set of processes $T$, we say that a view $\mathbb{V}$ is \emph{$T$-resilient} if and only if for every process $p_i\not\in T$, either $\mathbb{V}[i]=\bot$ or $\mathbb{V}[i]=\mathbb{F}[i]$.
\end{definition}

Intuitively, a correct process $p_i$ will either not have heard from $p_j\not\in A$ or it will have the correct assumption for $p_j$. Thus, $p_i$'s view is $A$-resilient at all times in execution $e$.

As we said, processes in $A$ may lie about their assumptions causing quorums to contain unreliable slices.
Moreover, processes in $A$ may aim at preventing intersection among quorums of correct processes.
In the following definition we characterize the notion of worst-case view, i.e., when faulty processes gossip only empty configurations.
By doing so, quorums of correct processes will contain fewer members, increasing the chances of an empty intersection among them.

\begin{definition}[Worst-case view]
  Given a set of processes $T$, the \emph{worst-case view} with respect to $T$ is the view $\mathbb{V}^*_T$ such that:
  \begin{enumerate}
    \item for every $p_i\in \Pi\setminus T$, $\mathbb{V}^*_T[i]=\mathbb{F}[i]$, and
    \item for every $p_i\in T$, $\mathbb{V}^*_T[i]=(\emptyset, \left\{\emptyset\right\})$.
  \end{enumerate}
\end{definition}

Finally, every quorum for a process $p_i \not\in A$ in a $A$-resilient view contains a quorum for $p_i$ in a worst-case view with respect to $A$. 
This is shown in the following lemma.

\begin{lemma}
  \label{lem:worst-case-quorums}
  Consider a set of processes $T$, a $T$-resilient view $\mathbb{V}$, and a process $p_i\not\in T$. Moreover, let us assume that processes in $T$ may lie about their assumptions.
  For every quorum $Q_i\in\mathscr{Q}(p_i,\mathbb{V})$, there exists a quorum $Q_i'\in\mathscr{Q}(p_i,\mathbb{V}^*_T)$ such that $Q_i'\subseteq Q_i$.
\end{lemma}
\begin{proof}
Let $T$ be a set of processes, $\mathbb{V}$ be a $T$-resilient view, $p_i$ a process not in $T$.
Since $\mathbb{V}$ is a $T$-resilient view, for every $p_j \not\in T$ it holds either $\mathbb{V}[j]=\bot$ or $\mathbb{V}[j]=\mathbb{F}[j]$.
However, processes in $T$ may lie about their assumptions and, because of that, the view of process $p_i \not\in T$ may contain arbitrary configurations received from processes in $T$.

If $Q_i\in\mathscr{Q}(p_i,\mathbb{V})$ is a quorum for $p_i$ in $\mathbb{V}$, then $Q_i$ might contain slices of processes in $T$ which are derived from false assumptions.
One can easily show that by starting from a $T$-compatible view and by removing the configurations received by processes in $T$, it is possible to obtain the corresponding worst-case view.
By removing configurations from $\mathbb{V}$, also $Q_i$ becomes smaller, i.e., with less members, obtaining a quorum $Q_i' \subseteq Q_i$.
In fact, by removing from $Q_i$ a slice $\overline{F}_j$ of a process $p_j \in T$, also slices of other processes in $\overline{F}_j$ might get removed in order for Definition~\ref{def:quorum-function} to be satisfied on $Q_i'$.
This proves the lemma.
\end{proof}

\section{Leagues}
\label{sec:leagues}

We now define the notion of a \emph{league}. In Section~\ref{sec:reliable-broadcast} we show how a league allows to implement Bracha broadcast. 


\begin{definition}[League]
\label{def:league}
 set of processes $L$ is a \emph{league} for the quorum function $\mathscr{Q}$ if and only if the following property holds:
  \begin{description}
  \item[Consistency:] For every set $T\subseteq \Pi$ tolerated by $L$, for every two $T$-resilient views $\mathbb{V}$ and $\mathbb{V}'$, for every two processes $p_i,p_j\in L\setminus T$, and for every two quorums $Q_i\in \mathscr{Q}(p_i,\mathbb{V})$ and $Q_j\in \mathscr{Q}(p_j,\mathbb{V}')$ it holds $\left( Q_i\cap Q_j \right) \setminus T \neq \emptyset$.
    \item[Availability:] For every set $T\subseteq \Pi$ tolerated by $L$ and for every $p_i \in L\setminus T$, there exists a quorum $Q_i \in \mathscr{Q}(p_i, \mathbb{F})$ for $p_i$ such that $Q_i \subseteq L \setminus T$. 
  \end{description}
\end{definition}

If we consider an execution $e$ tolerated by a league $L$, where $A$ is the set of faulty processes, the consistency property of $L$ implies that, at any time, any two quorums of correct processes in $L$ have some correct process in common. This is similar to the consistency property of symmetric and asymmetric Byzantine quorum systems~\cite{DBLP:journals/dc/MalkhiR98,DBLP:conf/opodis/CachinT19}.

Moreover, since the set of faulty processes $A$ is tolerated by $L$, by the availability property of $L$, every correct process in $L$ has a quorum in $\mathbb{F}$ consisting of only correct processes.

\begin{example}
\label{ex:league}
Observe that the set $\Pi$ as introduced in Example~\ref{ex:tolerated} is a league. 
In fact, for every set $T$ tolerated by $\Pi$, i.e., $\emptyset$, $\{p_1\}, \{p_4\}$ and $\{p_1, p_4\}$, for every two processes $p_i, p_j \in \Pi \setminus T$ and for every two quorums $Q_i \in \mathcal{S}_i$ and $Q_j \in \mathcal{S}_j$ as in Example~\ref{ex:quorums}, it holds $(Q_i \cap Q_j) \setminus T \neq \emptyset$, and for every $p_i \in \Pi \setminus T$, there exists a quorum $Q_i \in \mathcal{S}_i$ such that $Q_i \subseteq \Pi \setminus T.$
\end{example}

The following lemma shows that the union of two intersecting leagues $L_1$ and $L_2$ is again a league, assuming that for every set $T$ tolerated by both the leagues, $L_1$ and $L_2$ have a common process not in $T$.

\begin{lemma}
\label{lem:two-leagues}
  If $L_1$ and $L_2$ are two leagues such that $L_1 \cap L_2 \neq \emptyset$ and such that for every set $T$ tolerated by $L_1 \cup L_2$, there exists a process $p_k \in \left( L_1 \cap L_2\right) \setminus T$, then $L_1\cup L_2$ is a league.
\end{lemma}

\begin{proof}
Let $L_1$ and $L_2$ be two leagues such that $L_1 \cap L_2 \neq \emptyset$.
For every $T$ tolerated by $L_1 \cup L_2$ (and so, tolerated by $L_1$ and $L_2$, independently), for every $p_i\in L_1\setminus T$ and $p_j\in L_2\setminus T$, for every two $T$-resilient views $\mathbb{V}$ and $\mathbb{V}'$ for $p_i$ and $p_j$, respectively, let $Q_i\in\mathscr{Q}(p_i,\mathbb{V})$ and $Q_j\in\mathscr{Q}(p_j,\mathbb{V}')$ be two quorums for $p_i$ and $p_j$, respectively.
Observe that, by assumption, for every tolerated set $T$ by $L_1 \cup L_2$ there exists a process $p_k \in \left( L_1 \cap L_2\right) \setminus T$.
Let $p_k \in L_1 \cap L_2$ and let $Q_k \in \mathscr{Q}(p_k,\mathbb{V})$ be a quorum for $p_k$ such that $Q_k \subseteq L_1$, according to availability property of $L_1$.
From consistency property of $L_2$, $(Q_k\cap Q_j)\setminus T\neq \emptyset$ and every process in this intersection belongs to $L_1$.
Observe that, $Q_j$ is a quorum for every of its member. This implies that $Q_j$ is a quorum for every process in $(Q_k\cap Q_j)\setminus T$ and every process in $(Q_k\cap Q_j)\setminus T$ has quorum in $L_1$.
Moreover, $(Q_k\cap Q_i)\setminus T\neq \emptyset$.
It follows that $(Q_i \cap Q_j) \setminus T \neq \emptyset.$

Finally, by availability property of $L_1$ and $L_2$, for every tolerated set $T$ by $L_1$ and $L_2$ and for every process $p_i \in L_1 \setminus T$ and $p_j \in L_2 \setminus T$, eventually there exists a quorum $Q_i \in \mathscr(p_i, \mathbb{F})$ for $p_i$ and a quorum $Q_j \in \mathscr{Q}(p_j, \mathbb{F})$ for $p_j$ such that $Q_i \subseteq L_1 \setminus T$ and $Q_j \subseteq L_2 \setminus T$, respectively.
If $p_i = p_j \in L_1 \cap L_2$, then there exists a quorum $Q_i$ for $p_i$ such that $Q_i \subseteq (L_1 \cup L_2) \setminus T$.
\end{proof}

In the following lemma we show that we can characterize the consistency property of a league just by considering worst-case views. Intuitively, this result relies on the observation that every $T$-resilient view can be seen as extensions of worst-case views with respect to $T \subseteq \Pi$, in the sense that a $T$-resilient view can be obtained by starting from a worst-case view with respect to $T$ and by considering the non-empty configurations received by processes in~$T$.

\begin{lemma}
The consistency property a league $L$ holds if and only if for every set $T\subseteq \Pi$ tolerated by $L$, for every two worst-case views $\mathbb{V}^*_T$ and $\mathbb{V}'^*_T$ with respect to $T$, for every two processes $p_i,p_j\in L\setminus T$, and for every two quorums $Q_i\in \mathscr{Q}(p_i,\mathbb{V}^*_T)$ and $Q_j\in \mathscr{Q}(p_j,\mathbb{V}'^*_T)$ it holds $\left( Q_i\cap Q_j \right) \setminus T \neq \emptyset$.
\end{lemma}

\begin{proof}
Let us assume that the consistency property of a league $L$ holds. Since the property holds for every pair of views, it must hold also for worst-case views. The implication easily follows.

Let us now assume that for every set $T\subseteq \Pi$ tolerated by $L$, for every two worst-case views $\mathbb{V}^*_T$ and $\mathbb{V}'^*_T$ with respect to $T$, for every two processes $p_i,p_j\in L\setminus T$, and for every two quorums $Q_i\in \mathscr{Q}(p_i,\mathbb{V}^*_T)$ and $Q_j\in \mathscr{Q}(p_j,\mathbb{V}'^*_T)$ it holds $\left( Q_i\cap Q_j \right) \setminus T \neq \emptyset$.
Observe that, given a quorum $Q_i \in \mathscr{Q}(p_i, \mathbb{V}^*_T)$ for $p_i$ in a worst-case view $\mathbb{V}^*_T$, all the quorums obtained by also considering all the possible configurations received from processes in $T$ that are not in $\mathbb{V}^*_T$ do contain $Q_i$.
Moreover, there cannot exist a $T$-resilient view that does not consist of configurations of a worst-case view with respect to $T$.
If this was the case, then by removing configurations received from processes in $T$ one would obtain a worst-case view with respect to $T$, reaching a contradiction.
So, all the quorums obtained from $T$-resilient views will also intersect in processes that are not contained in~$T$.
\end{proof}

Now we show how a league can be abstracted and defined without considering views.
This will be useful in Section~\ref{sec:comparison} when we compare our model with other permissionless models.
First, we introduce the following definitions.

\begin{definition}[Inclusive up to]
\label{def:transitive}
  A set $I \subseteq \Pi$ is \emph{inclusive up to} a set $T \subseteq \Pi$ if and only if for every $p_i \in I \setminus T$, process $p_i$ has a slice in $I$.
\end{definition}

If we consider an execution $e$ with set of faulty processes $A$ then a set of processes $I$ is inclusive up to $A$ if and only if every correct process in $I$ has a slice contained in $I$. 

\begin{definition}[Rooted at]
\label{def:rooted}
  A set $R \subseteq Pi$ is \emph{rooted at a process $p_i$} if and only if $p_i$ has a slice in $R$.
  A set $R \subseteq \Pi$ is \emph{rooted in a set} $T' \subseteq \Pi$ whenever $R$ is rooted at a member of $T'$.
\end{definition} 

\begin{lemma}
\label{lem:quorum-transit}
If $\mathbb{V}$ is a $T$-resilient view and $Q_i \in \mathcal{Q}(p_i,\mathbb{V})$ for some process $p_i$, then $Q_i $ is inclusive up to $T$ and rooted at $p_i$.
\end{lemma}

\begin{proof}
If $\mathbb{V}$ is a $T$-resilient view then, by Definition~\ref{def:S-resilient}, processes outside $T$ do not lie about their assumptions.
  By definition of a quorum $Q_i$ for a process $p_i$ in a view $\mathbb{V}$, every process in $Q_i$, and so in $Q_i \setminus T$, has a slice in $Q_i$.
  This implies that $Q_i$ is inclusive up to $T$ and rooted at $p_i$.
\end{proof}

In the following lemma we show that given a set of processes $T$ tolerated by $L\subseteq \Pi$, for every set of processes $I$ inclusive up to $T$ and rooted at $p_i \in L\setminus T$ it is possible to find a $T$-resilient view in which $I$ is a quorum for $p_i$. This view is a worst-case view with respect to~$T$.

\begin{lemma}
\label{lem:transit-quorum}
Let $L$ be a set of processes. For every set $T \subseteq \Pi$ tolerated by $L$, if $I \subseteq \Pi$ is a set inclusive up to $T$ and rooted at $p_i \in L\setminus T$, then there is a $T$-resilient view in which $I$ is a quorum for $p_i$.
\end{lemma}

\begin{proof}
Let $T \subseteq \Pi$ be a tolerated set by a set of processes $L$ and let $I \subseteq \Pi$ be a set inclusive up to $T$ and rooted at $p_i \in L \setminus T$.
This implies that $p_i$ and every other process $p_j \in I\setminus T$ have a slice in $I$.
 Let us consider the worst-case view $\mathbb{V}^*_T$ with respect to $T$. Clearly, $\mathbb{V}^*_T$ is $T$-resilient. This implies that, in $\mathbb{V}^*_T$, the set $I$ is a quorum for $p_i$, i.e., $I \in \mathscr{Q}(p_i, \mathbb{V}^*_T)$.
\end{proof}

\begin{remark}
\label{remark:inclusive}
Observe that given a set of processes $L$ and a worst-case view $\mathbb{V}^*_T$ with respect to a set $T\subseteq \Pi$ tolerated by $L$, every quorum $Q_i \in \mathscr{Q}(p_i, \mathbb{V}^*_T)$ for $p_i \in L\setminus T$ is inclusive up to $T$ and rooted at $p_i$.

Moreover, given the set $\mathcal{I}$ of all the sets $I\subseteq \Pi$ inclusive up to $T$ and rooted at $p_i \in L\setminus T$, the set $\mathcal{I}$ contains all the quorums $Q_i \in \mathscr{Q}(p_i, \mathbb{V})$ for every $T$-resilient view $\mathbb{V}$. 
In  fact, by Definition~\ref{def:transitive}, given a set of processes $I$ inclusive up to a set of processes $T$, the requirement of having a slice in $I$ is only for processes in $I\setminus T$, leaving processes in $T \cap I$ with no requirements on the choice of their slices.

However, given a $T$-resilient view $\mathbb{V}$, by Definition~\ref{def:quorum-function}, a quorum $Q_i$ for $p_i$ requires instead every process in $Q_i$ to have a slice contained in $Q_i$.
This means that given a $T$-resilient view $\mathbb{V}$, quorum $Q_i$ for $p_i$ is contained in $\mathcal{I}$, being a special case of inclusive set up to $T$.
\end{remark}

\begin{lemma}
\label{lem:consistency}
The consistency property of a league $L$ holds if and only if for every set $T \subseteq \Pi$ tolerated by $L$, and for every two sets $I\subseteq \Pi$ and $I'\subseteq \Pi$ that are rooted at $L \setminus T$ and inclusive up to $T$ it holds $\left( I\cap I'\right) \setminus T \neq \emptyset$.
\end{lemma}

\begin{proof}

 Let us assume that the consistency property of a league $L$ holds.
 Suppose by contradiction that there is a set $T \subseteq \Pi$ tolerated by $L$ and two sets $I\subseteq \Pi$ and $I'\subseteq \Pi$ that are inclusive up to $T$ and rooted at $L\setminus T$ in $p_i$ and $p_j$, respectively, such that $\left( I \cap I' \right) \setminus T = \emptyset$.

 By Lemma~\ref{lem:transit-quorum}, there are a $T$-resilient view $\mathbb{V}$ in which $I$ is a quorum for $p_i$ and a $T$-resilient view $\mathbb{V}'$ in which $I'$ is a quorum for $p_j$ and we reached a contradiction.
 
Let us now assume that for every set $T \subseteq \Pi$ tolerated by $L$, and for every two sets $I\subseteq \Pi$ and $I'\subseteq \Pi$ that are inclusive up to $T$ and rooted at $L \setminus T$ it holds $\left( I\cap I'\right) \setminus T \neq \emptyset$. 

Let $\mathcal{I}$ and $\mathcal{I}'$ be the sets of all the sets $I \subseteq \Pi$ and $I' \subseteq \Pi$ inclusive up to $T$ and rooted at $p_i \in L\setminus T$ and $p_j \in L\setminus T$, respectively. The proof follows from the reasoning in Remark~\ref{remark:inclusive}: for every two $T$-resilient views $\mathbb{V}$ and $\mathbb{V}'$, every quorum $Q_i \in \mathscr{Q}(p_i, \mathbb{V})$ for $p_i$ is contained in $\mathcal{I}$ and every quorum $Q_j \in \mathscr{Q}(p_j, \mathbb{V})$ for $p_j$ is contained in $\mathcal{I}'$.
\end{proof}

\begin{lemma}
\label{lem:availability}
The availability property of a league $L$ holds if and only if for every set $T\subseteq \Pi$ tolerated by $L$, every member of $L\setminus T$ has a survivor set in $L \setminus T$.
\end{lemma}

\begin{proof}

Let us assume that the availability property of a league $L$ holds, i.e., for every set of processes $T$ tolerated by $L$ and for every $p_i \in L\setminus T$, there exists a quorum $Q_i \in \mathscr{Q}(p_i, \mathbb{F})$ for $p_i$ such that $Q_i \subseteq L \setminus T$. 
This means that, by Definition~\ref{def:survset}, every process in $L \setminus S$ has a survivor set in $L\setminus S$.

Let us now assume that for every set $T\subseteq \Pi$ tolerated by $L$, every member of $L\setminus T$ has a survivor set $S$ in $L \setminus T$.
Let $p_i$ be a process in $L$, by Definition~\ref{def:quorum-function} we have that $S \in \mathscr{Q}(p_i, \mathbb{F})$ for $p_i$. The proof follows.
\end{proof}

\section{Comparison with Other Models}
\label{sec:comparison}

In this section we compare our model with the classic model based on fail-prone systems~\cite{DBLP:journals/dc/MalkhiR98}, with the asymmetric model\cite{DBLP:conf/asiacrypt/DamgardDFN07, DBLP:conf/opodis/CachinT19}, with the federated Byzantine agreement system model~\cite{Mazieres2015TheSC}, and with the personal Byzantine quorum system model~\cite{DBLP:conf/wdag/LosaGM19}.

\subsection{Comparison with Fail-Prone Systems}

We show that classic fail-prone systems and quorums can be understood as a special case of our model, when every process knows the entire system and assumes the same, global fail-prone system.

Let $\Pi=\{p_1,\ldots,p_n\}$ be a set of processes. A Byzantine quorum system for a fail-prone system $\mathcal{F}$ (Definition~\ref{def:fail-prone-system}) satisfies (Consistency) $\forall Q_1, Q_2 \in \mathcal{Q} , \forall F \in \mathcal{F}: \, Q_1 \cap Q_2 \not
    \subseteq F$; and (Availability) $\forall F \in \mathcal{F}: \, \exists~Q \in \mathcal{Q}: \, F \cap Q = \emptyset.$ Moreover, a Byzantine quorum system $\mathcal{Q}$ for $\mathcal{F}$ exists if and only if the $Q^3$-condition holds~\cite{DBLP:journals/dc/MalkhiR98,DBLP:journals/joc/HirtM00}, which means that for every $F_1, F_2, F_3 \in \mathcal{F}: \, \Pi \not\subseteq F_1 \cup F_2 \cup F_3$.
 In particular, if $Q^3(\mathcal{F})$ holds, then
  $\overline{\mathcal{F}}$, the bijective complement of $\mathcal{F}$, is a Byzantine quorum system; this is the Byzantine quorum system consisting of survivor sets~\cite{DBLP:conf/icdcs/JunqueiraM03}. Notice that, in a classic system, the failures that are tolerated by the processes are all possible subsets of fail-prone sets in $\mathcal{F}$ and we have $P_i = \Pi$ for every $p_i \in \Pi$. Every process also knows the global quorum system.

  We define a bijective function $f$ between the set of fail-prone systems and a subset of PFPS such that $f(\mathcal{F}) = [(\Pi,\mathcal{F}),\ldots,(\Pi,\mathcal{F})]$ with $n$ repetitions and we notice that in classic fail-prone systems there is only one view, namely $\mathbb{V} = f(\mathcal{F})$.

We define the quorum function $\mathscr{Q}: \Pi \times \Upsilon \rightarrow 2^{\Pi}$ such that for every process $p_i \in \Pi$, $\mathscr{Q}(p_i,f(\mathcal{F})) = \overline{\mathcal{F}}$.  Observe that any set in $\overline{\mathcal{F}}$ is a slice of every process $p_i \in \Pi$ according to Definition~\ref{def:slice}. In the next theorem we consider this quorum function and show that, given some assumptions on $\mathcal{F}$, any set in $\overline{\mathcal{F}}$ is also a quorum for every process $p_i \in \Pi$ according to Definition~\ref{def:quorum-function}.

\begin{theorem}
\label{thm:q3-league}
Let $\Pi$ be the set of all processes and $\mathcal{F}$ the fail-prone system for $\Pi$.
Then $Q^3(\mathcal{F})$ holds if and only if $\Pi$ is a league for the quorum function $\mathscr{Q}$ in $f(\mathcal{F})$.

\end{theorem}


\begin{proof}
Let us first assume that $Q^3(\mathcal{F})$ holds.
This means that for every $F_1, F_2, F_3 \in \mathcal{F}$, $ \Pi \not\subseteq F_1\cup F_2\cup F_3$. It follows that $\overline{\mathcal{F}}$ is a quorum system for $\mathcal{F}$. 
Consistency property of $\overline{\mathcal{F}}$ implies that for every tolerated set $F$, for every two processes $p_i$ and $p_j$ in $\Pi \setminus F$ and for every two quorums $Q_i \in \mathscr{Q}(p_i, f(\mathcal{F}))$ and $Q_j \in \mathscr{Q}(p_j, f(\mathcal{F}))$ for $p_i$ and $p_j$, respectively, it holds that $(Q_i \cap Q_j) \setminus T \neq \emptyset.$

 The availability property of $\overline{\mathcal{F}}$ implies that for every set $F \in \mathcal{F}$ tolerated by $\Pi$, every process $p_i \in \Pi \setminus F$ has a quorum in $\Pi \setminus F$: given $F$, there exists a quorum $Q \in \mathscr{Q}(p_i, f(\mathcal{F}))$ such that $Q \cap F = \emptyset$ and $Q \subseteq \Pi \setminus F$. It follows that $\Pi$ is a league for the quorum function $\mathscr{Q}$ in $f(\mathcal{F})$.

Let us now assume that $\Pi$ is a league for the quorum function $\mathscr{Q}$ in $f(\mathcal{F})$.
The consistency property of $\Pi$ implies that for every $T$ tolerated by $\Pi$ (which are all the sets in $\mathcal{F}$), for every two processes $p_i$ and $p_j$ in $\Pi \setminus T$, for every two quorums $Q_i \in \overline{\mathcal{F}}$ and $Q_j \in \overline{\mathcal{F}}$ for $p_i$ and $p_j$, respectively, it holds $(Q_i \cap Q_j) \setminus T \neq \emptyset.$
Moreover, by availability property of $\Pi$ there exists a quorum in $\Pi \setminus T$ (which is the same for every process $p_i \not\in T$). This implies that, for every fail-prone set $F \in \mathcal{F}$, there is a quorum $Q_i$ such that $Q_i \cap F = \emptyset$.

These two facts imply that $\overline{\mathcal{F}}$ is a classic Byzantine quorum system for $\mathcal{F}$ and so $Q^3(\mathcal{F})$ holds.
\end{proof}

\subsection{Comparison with Asymmetric Fail-Prone Systems}

In the asymmetric model~\cite{DBLP:conf/asiacrypt/DamgardDFN07}, every process is free to express its own trust assumption about the processes in one common globally known system through a subjective fail-prone system.

An asymmetric fail-prone system $\mathbb{F}' = [\mathcal{F}'_1, \dots, \mathcal{F}'_n]$ consists of an array of fail-prone systems, where $\mathcal{F}'_i \subseteq 2^{\Pi}$ denotes the trust assumption of $p_i$.

An \emph{asymmetric Byzantine quorum system} for $\mathbb{F}'$~\cite{DBLP:conf/opodis/CachinT19}, is an array of
  collections of sets $\mathbb{Q}' = [\mathcal{Q}'_1, \dots, \mathcal{Q}'_n]$, where
  $\mathcal{Q}'_i \subseteq 2^{\Pi}$ for $i \in [1,n]$.  The set
  $\mathcal{Q}'_i \subseteq 2^{\Pi}$ is called the \emph{quorum system of $p_i$} and
  any set $Q_i \in \mathcal{Q}'_i$ is called a \emph{quorum (set) for $p_i$}.
  Moreover, defining $\mathcal{F}'^*= \{ F' | F' \subseteq F, F \in \mathcal{F}' \}$, the following conditions hold: (Consistency) $\forall i,j \in [1,n],~\forall Q_i \in \mathcal{Q}'_i, \forall Q_j \in \mathcal{Q}'_j, \forall F_{ij} \in {\mathcal{F}'_i}^* \cap {\mathcal{F}'_j}^*: \,
      Q_i \cap Q_j \not\subseteq F_{ij}$; and (Availability) $\forall i \in [1,n], \forall F_i \in \mathcal{F}'_i: \, \exists~Q_i \in \mathcal{Q}'_i: \, F_i \cap Q_i = \emptyset.$
  
The $Q^3$-condition in the classic model can be generalized as follows: we say that $\mathbb{F}'$ satisfies the \emph{$B^3$-condition}~\cite{DBLP:conf/asiacrypt/DamgardDFN07}, abbreviated as $B^3(\mathbb{F}')$, whenever it holds for all $i,j \in [1,n]$ that $\forall F_i \in \mathcal{F}'_i, \forall F_j\in\mathcal{F}'_j,
    \forall F_{ij} \in {\mathcal{F}'_i}^*\cap{\mathcal{F}'_j}^*: \,
    \Pi \not\subseteq F_i \cup F_j \cup F_{ij}.$

An asymmetric fail-prone system $\mathbb{F}'$ satisfying the $B^3$-condition is sufficient and necessary~\cite{DBLP:conf/opodis/CachinT19} for the existence of a corresponding asymmetric quorum system $\mathbb{Q}'=[\mathcal{Q}'_1,\ldots,\mathcal{Q}'_n]$, with $\mathcal{Q}'_i = \overline{\mathcal{F}'}_i$. Processes in this model are classified in three different types, given an execution $e$ with faulty set $A$: a process $p_i \in A$ is \emph{faulty}, a correct process $p_i$ for which $A \not\in {\mathcal{F}'_i}^*$
  is called \emph{naive}, while a correct process $p_i$ for which $A \in {\mathcal{F}'_i}^*$ is called \emph{wise}.

Finally, a \emph{guild} $\mathcal{G}$ for $A$ is a set of wise processes that contains at least one quorum for each member.

Let $\Pi$ be a set of processes in the asymmetric model and $\mathbb{F}'=[\mathcal{F}'_1,\ldots,\mathcal{F}'_n]$ be an asymmetric fail-prone system.
Define the function $g$ from asymmetric fail-prone systems to PFPS such that $g(\mathbb{F}') = [(\Pi,\mathcal{F}'_1),\ldots,(\Pi,\mathcal{F}'_n)]$.
Observe that, in an asymmetric system, the failures that may be tolerated by the processes are possible subsets of fail-prone sets in the fail-prone systems of $\mathbb{F}'$ and $P_i = \Pi$ for every $p_i \in \Pi$. 
Moreover, as in the classic model, there is only one view, which is $\mathbb{V} = g(\mathbb{F}')$.

We define the quorum function $\mathscr{Q}: \Pi \times \Upsilon \rightarrow 2^{\Pi}$ such that for every guild $\mathcal{G} \subseteq 2^{\Pi}$, if $p_i \in \mathcal{G}$ then $\mathscr{Q}(p_i,g(\mathbb{F}')) = \{\mathcal{G}, \Pi\}$, otherwise $\mathscr{Q}(p_i,g(\mathbb{F}')) = \{\Pi\}$.

Observe that a quorum in the asymmetric model is a slice according to Definition~\ref{def:slice} and, for every process $p_i \in \Pi$, every set in $\mathscr{Q}(p_i,g(\mathbb{F}'))$ is a quorum according to Definition~\ref{def:quorum-function}.

Through the following theorem we establish the relationship between the asymmetric model and the permissionless model.

\begin{theorem}
\label{thm:b3-implies-perm-b3}
Let us consider an asymmetric model among a set $\Pi$ of processes with asymmetric fail-prone system $\mathbb{F}'$.
If $B^3(\mathbb{F}')$ holds and $\Pi$ tolerates some sets $T \subseteq \Pi$, then there exists a quorum function $\mathscr{Q}$ such that $\Pi$ is a league in $g(\mathbb{F}')$.
\end{theorem}

\begin{proof}
 Let us assume that $\Pi$ tolerates some sets $T \subseteq \Pi$ and let us consider the quorum function $\mathscr{Q}$ define in this section in the context of the asymmetric model.
 This means that, for every set $T$ tolerated by $\Pi$, every process $p_i \in \Pi\setminus T$ has a slice contained in $\Pi \setminus T$.
 This implies that in every execution in which $T$ is the set of faulty processes, every process in $\Pi \setminus T$ is wise and $\Pi \setminus T$ is a guild.
 
 Moreover, let us also assume that $B^3(\mathbb{F}')$ holds. This implies the existence of an asymmetric Byzantine quorum system $\mathbb{Q}'$ such that for every set $T$ tolerated by $\Pi$, for every two processes $p_i$ and $p_j$ in $\Pi \setminus T$ and for every two quorums $Q_i \in \mathcal{Q}'_i$ and $Q_j \in \mathcal{Q}'_j$ for $p_i$ and $p_j$, respectively, it holds that $\left( Q_i \cap Q_j \right)  \setminus T \neq \emptyset.$ Observe that the set $\Pi \setminus T \in \mathscr{Q}(p_i, g(\mathbb{F}'))$ is a quorum in the permissionless model for every $p_i \in \Pi \setminus T$ according to Definition~\ref{def:quorum-function}. This implies that $\Pi$ satisfies availability property of a league.
 
Finally, for the consistency property observe that for every process $p_i\in \Pi$, the set system $\mathscr{Q}(p_i, g(\mathbb{F}'))$ satisfies Definition~\ref{def:quorum-function}; by construction we have at most only two quorums for every $p_i$ which are $\Pi$ and $\Pi\setminus T$ both satisfying Definition~\ref{def:quorum-function}. Consistency of $\mathbb{Q}'$ implies intersection among the quorums in $\mathscr{Q}(p_i, g(\mathbb{F}'))$, for every process in $\Pi\setminus T$.

It follows that $\Pi$ is a league for the quorum function $\mathscr{Q}$ in $g(\mathbb{F}')$.
\end{proof}

Theorem~\ref{thm:b3-implies-perm-b3} shows a relation between the asymmetric model and the permissionless model. 
In particular, if $B^3(\mathbb{F}')$ holds and $\Pi$ tolerates some sets $T$, then the quorum function $\mathscr{Q}$ makes $\Pi$ a league. However, we could have scenarios in which only a subset of $\Pi$ tolerates some sets $T$. 
In particular, we have the following result.

\begin{lemma}
Let $\Pi=\{p_1,\ldots,p_n\}$ be a set of processes, $\mathbb{F}'$ be an asymmetric fail-prone system over $\Pi$ and $g(\mathbb{F}')$ the corresponding PFPS as described in the text.
Moreover, let us consider an execution $e$ with set of faulty processes $A$ with guild $\mathcal{G}$.
Then, $\mathcal{G}$ is the only set that tolerates $e$.
\end{lemma}

\begin{proof}
By definition of guild, every process in $\mathcal{G}$ is wise and has a quorum contained in $\mathcal{G}$.
Observe that, given a wise process $p_i$, there exists a fail-prone set $F \in \mathcal{F}'_i$ in $\mathbb{F}'$ such that $A \subseteq F$.
Moreover, a quorum $Q_i$ for $p_i$ in the asymmetric model satisfies Definition~\ref{def:slice} and it is then a slice of $p_i$.
This implies that every process in $\mathcal{G}$ has its assumptions satisfied according to Definition~\ref{def:assms-satisfied}.
Moreover, every process in $\mathcal{G}$ has a slice contained in $\mathcal{G}$. 
\end{proof}

In the following lemma we characterize a link between the notion of a guild, in a given execution, and a league.

\begin{lemma}
Let us consider an asymmetric Byzantine quorum system $\mathbb{Q}'$ and a guild $\mathcal{G}$ in any execution with set of faulty processes $A$. 
Then, $\mathcal{G}$ is a league for the quorum function $\mathscr{Q}$ in $g(\mathbb{F}')$.
\end{lemma}

\begin{proof}
The result follows from Theorem~\ref{thm:b3-implies-perm-b3} by applying the same reasoning with $\mathcal{G}$ instead of $\Pi \setminus T$ as a guild.
\end{proof}

In the following lemma we show a scenario where no asymmetric Byzantine quorum systems exist but it is possible to find a league for $\mathscr{Q}$ in $g(\mathbb{F}')$.

\begin{lemma}
  There exists an asymmetric fail-prone system $\mathbb{F}'$ such that:
  \begin{itemize}
    \item there is no asymmetric Byzantine quorum system for $\mathbb{F}'$, but
    \item there exists a quorum function $\mathscr{Q}$ that make $\Pi$ a league in $g(\mathbb{F}')$.
  \end{itemize}
\end{lemma}

\begin{proof}
We prove this lemma through an example with four processes.
  Consider an asymmetric fail-prone system $\mathbb{F}'_4$ over four processes $p_1$, $p_2$, $p_3$, and $p_4$ with $\mathcal{F}'_1=\{\{p_3,p_4\}\}$, $\mathcal{F}'_2=\{\{p_1,p_4\}\}$, $\mathcal{F}'_3=\{\{p_1,p_4\}\}$, and $\mathcal{F}'_4=\{\{p_1,p_2\}\}$, as in Example~\ref{ex:tolerated}.

  Observe that, by the availability property of an asymmetric Byzantine quorum system, $p_1$ must have a quorum in $\{p_1,p_2\}$ and $p_4$ must have a quorum in $\{p_3,p_4\}$. Since $\{p_1,p_2\}$ and $\{p_3,p_4\}$ are disjoint, it is impossible to satisfy the consistency property. Thus, there does not exist any asymmetric Byzantine quorum system for $\mathbb{F'}_4$.
  Another way to see this is by observing that $B^3(\mathbb{F}'_4)$ does not hold: $\{p_3,p_4\} \cup \{p_1,p_2\} = \Pi.$

  However, as shown in Example~\ref{ex:league}, $\Pi$ is a league in $g(\mathbb{F}'_4)$. So, there is no asymmetric Byzantine quorum system for $\mathbb{F}'_4$ but $\mathscr{Q}$ makes $\Pi$ a league in~$g(\mathbb{F}'_4)$. 
\end{proof}

\subsection{Comparison with Federated Byzantine Agreement Systems}

The federated Byantine agreement system (FBAS) model has been introduced by Mazi{\`e}res~\cite{Mazieres2015TheSC} in the context of the Stellar white paper. 
Differently from the models presented before in this section, the FBAS model is a permissionless model, where processes, each with an initial set of known processes, continuously discover new processes. 
In a FBAS, every process $p_i$ chooses a set of slices, which are sets of processes sufficient to convince $p_i$ of agreement and a set of processes $Q_i$ is a quorum for $p_i$ whenever $p_i$ has at least one slice inside $Q_i$ and every member of $Q_i$ has a slice that is a subset of $Q_i$. 
In particular, a quorum $Q_i$ is a quorum for every of its members. However, despite the permissionless nature of a FBAS, a global intersection property among quorums is required for the analysis of the Stellar Consensus Protocol (SCP), and the scenario with disjoint quorums is not considered by Mazi{\`e}res.

A central notion in FBAS is that of \emph{intact} set; given a set of processes $\Pi$, an execution with set of faulty processes $A$ and a set of correct processes $\mathcal{W} = \Pi \setminus A$, a set of processes $\mathcal{I} \subseteq \mathcal{W}$ is an \emph{intact set}~\cite{DBLP:conf/opodis/Garcia-PerezG18, DBLP:conf/wdag/LosaGM19}  when the following conditions hold: (Consistency) for every two processes $p_i$ and $p_j$ in $\mathcal{I}$ and for every two quorums $Q_i$ and $Q_j$ for $p_i$ and $p_j$, respectively, $Q_i \cap Q_j \cap \mathcal{I} \neq \emptyset$; and (Availability) $\mathcal{I}$ is a quorum for every of its members.

Every process in $\mathcal{I}$ is called \emph{intact}, while every process in $\Pi \setminus \mathcal{I}$ (correct or faulty) is called \emph{befouled} and some properties of the Stellar Consensus Protocol are guaranteed only for intact processes. Moreover, the union of two intersecting intact sets is an intact set.
Finally, by requiring a system-wide intersection among quorums (as in the case of the SCP) one obtains an unique intact set (Lemma 34,~\cite{DBLP:conf/opodis/Garcia-PerezG18}).

We first show that our model generalizes the FBAS model by showing that a quorum in FBAS satisfies Definition~\ref{def:quorum-function}.

In FBAS a notion of fail-prone system is missing and definitions are given with respect to an execution with a fixed set of faulty processes $A$. 
However, because processes define slices, an implicit fail-prone system for every process can be derived.

In particular, given a set of processes $\Pi = \{p_1, p_2, \ldots\}$, every process in $\Pi$ defines its slices based on a known subset $P_i \subseteq \Pi$ by $p_i$ and $S_i$ is a slice for $p_i \in \Pi$ if and only if $p_i \in S_i$ and $S_i \subseteq P_i$~\cite{DBLP:conf/opodis/Garcia-PerezG18}. 
Let $\mathcal{S}_i \subseteq 2^\Pi$ be the set of slices of $p_i$, we can derive the following definition.

\begin{definition}(Federated fail-prone system)
\label{def:fed-fail-prone-set}
A set $F \subseteq \Pi$ is a \emph{fail-prone set} of $p_i$ if and only if there exists a slice $S_i \in \mathcal{S}_i$ of $p_i$ such that $F = P_i \setminus S_i$.
The set $\mathcal{F}''_i\subseteq 2^\Pi$ of all the fail-prone sets of $p_i$ is called \emph{fail-prone system} of $p_i$.
Finally, we call the set $\mathbb{F}''=[(P_1, \mathcal{F}''_1),(P_2, \mathcal{F}''_2),\ldots ]$ the \emph{federated fail-prone system}.
\end{definition}

In a FBAS, processes discover other processes' slices during an execution and so $p_i$ implicitly learns other processes' federated fail-prone sets. Moreover, correct processes do not lie about their slices~\cite{Mazieres2015TheSC, DBLP:conf/opodis/Garcia-PerezG18}.

It easy to observe that given different sets of slices received from different processes, Definition~\ref{def:fed-fail-prone-set} implies Definition~\ref{def:view}, obtaining a notion of view $\mathbb{V}$ in the FBAS model, and, because correct processes do not lie about their slices, Definition~\ref{def:S-resilient}.
We define the set $\Upsilon'$ to be the set of all the possible views in the FBAS model.

Given the notion of view in the FBAS model, we define the quorum function $\mathscr{Q}: \Pi \times \Upsilon' \rightarrow 2^{\Pi}$ such that $\mathscr{Q}(p_i,\mathbb{V})$ contains all the sets $Q_i$, called quorums, with $p_i \in Q_i$ and such that every process $p_j\in Q_i$ has a slice in $Q_i$. 
So, a quorum as defined by Mazi{\`e}res~\cite{Mazieres2015TheSC} satisfies Definition~\ref{def:quorum-function}.
Finally, in the FBAS model we introduce the notion of survivor set as defined in Definition~\ref{def:survset}.

In the following theorem we show that, by assuming a stronger consistency property for a league $L$, i.e., that the intersection among any two quorums of any two correct processes in the league contains some correct member of the league, then $L$ is an intact set in every execution tolerated by $L$.

\begin{theorem}
\label{thm:league-intact}
Let $L$ be a league for the quorum function $\mathscr{Q}$ and let us assume that for every set $T\subseteq \Pi$ tolerated by $L$, for every two $T$-resilient views $\mathbb{V}$ and $\mathbb{V}'$, for every two processes $p_i,p_j\in L\setminus T$, and for every two quorums $Q_i\in \mathscr{Q}(p_i,\mathbb{V})$ and $Q_j\in \mathscr{Q}(p_j,\mathbb{V}')$ it holds $\left( Q_i\cap Q_j \cap L \right) \setminus T \neq \emptyset$, then $L$ is an intact set for every every set $T\subseteq \Pi$ tolerated by $L$.
\end{theorem}

\begin{proof}
Let $L$ be a league for the quorum function $\mathscr{Q}$ and let $T$ be a set of processes tolerated by $L$.
If for every two $T$-resilient views $\mathbb{V}$ and $\mathbb{V}'$, for every two processes $p_i,p_j\in L\setminus T$, and for every two quorums $Q_i\in \mathscr{Q}(p_i,\mathbb{V})$ and $Q_j\in \mathscr{Q}(p_j,\mathbb{V}')$ it holds $\left( Q_i\cap Q_j \cap L \right) \setminus T \neq \emptyset$, then the consistency property of intact sets follows.
The availability property of an intact set follows by observing that, in $\mathbb{F}$, the set $L\setminus T$ is a quorum for every of its members.
\end{proof}




Observe that without the stronger consistency property assumed in Theorem~\ref{thm:league-intact}, since quorums of correct processes in $L$ may intersect in correct processes (not necessarily in $L$), it may be the case that $L$ is not an intact set.

\subsection{Comparison with Personal Byzantine Quorum Systems}

The personal Byzantine quorum system (PBQS) model has been introduced by Losa \emph{et al.}~\cite{DBLP:conf/wdag/LosaGM19} in the context of Stellar consensus aiming at removing the system-wide intersection property among quorums required by Mazi{\`e}res~\cite{Mazieres2015TheSC} for the SCP.

In the PBQS model a quorum for $p_i$ is a non-empty set of processes $Q_i$ such that if $Q_i$ is a quorum for $p_i$ and $p_j \in Q_i$, then there exists a quorum $Q_j$ for $p_j$ such that $Q_j \subseteq Q_i.$
In other terms, a quorum $Q_i$ for some process $p_i$ must contain a quorum for every one of its members. Losa \emph{et al.} point out that a global consensus among processes may be impossible since the full system membership is not known by the processes, and define the notion of \emph{consensus cluster} as a set of processes that can instead solve a \emph{local} consensus, i.e., consensus among the processes in a consensus cluster can be solved. In particular, given an execution with set of faulty processes $A$, a set of correct processes $\mathcal{C}$ is a consensus cluster when the following conditions hold: (Consistency) for every two processes $p_i$ and $p_j$ in $\mathcal{C}$ and for every two quorums $Q_i$ and $Q_j$ for $p_i$ and $p_j$, respectively, $Q_i \cap Q_j \not\subseteq A$; and (Availability) for every $p_i \in \mathcal{C}$ there exists a quorum $Q_i$ for $p_i$ such that $Q_i \subseteq \mathcal{C}.$ Losa \emph{et al.} prove that the union of two intersecting consensus clusters is a consensus cluster and that maximal consensus clusters are disjoint.
The latter implies that maximal consensus clusters might diverge from each other.

In the following we show a relationship between the notions of league and consensus cluster.
To do so, we first show that a quorum $Q_i$ for $p_i$ as defined in Definition~\ref{def:quorum-function} is also a quorum for $p_i$ in the PBQS model.

\begin{lemma}
\label{lem:quorum-is-pbqs}
Let $Q_i \in \mathscr{Q}(p_i, \mathbb{V})$ be a quorum for a process $p_i$ in a view $\mathbb{V}$ according to Definition~\ref{def:quorum-function}. Then $Q_i$ is a quorum for $p_i$ in the PBQS model.
\end{lemma}

\begin{proof}
Definition~\ref{def:quorum-function} implies that $Q_i$ is a quorum for every of its members. This means that for every process $p_j \in Q_i$, the set $Q_i$ is a quorum for $p_j$ such that $Q_i \subseteq Q_i$.
The result follows.
\end{proof}

In the following result we show that, given a league $L$, for every set $T \subseteq \Pi$ tolerated by $L$, the set $L \setminus T$ is a consensus cluster.

\begin{theorem}
Let $L$ be a league for the quorum function $\mathscr{Q}$. Then, for every set $T \subseteq \Pi$ tolerated by $L$, the set $L \setminus T$ is a consensus cluster.
\end{theorem}

\begin{proof}
Let $L$ be a league for the quorum function $\mathscr{Q}$ and let $T$ be a set of processes tolerated by $L$.
Lemma~\ref{lem:quorum-is-pbqs} implies that for every process $p_i$ and for every view $\mathbb{V}$, all the quorums in $\mathscr{Q}(p_i,\mathbb{V})$ for $p_i$ are quorums in the PBQS model.
So, the consistency and availability properties of a league imply that $L \setminus T$ satisfies the consistency and availability properties of a consensus cluster, making $L \setminus T$ a consensus cluster.
\end{proof}

\section{Permissionless Shared Memory}
\label{sec:shared-memory}

In this section we present a first application of permissionless quorum systems by showing how to emulate shared memory, represented by a \emph{register}.

A {register} stores values and can be accesses through two operations: $\op{write}(v)$, parameterized by a value $v$ belonging to a domain $\mathcal{V}$, and outputs a token \str{ack} when it completes; and $\op{read}$, which takes no parameter and outputs a value $v \in \mathcal{V}$ upon completion.
In this work we consider only a \emph{single-writer} register, where only a designated process $p_w$ may invoke $\op{write}$,
and allow \emph{multiple readers}, i.e., every process may execute a read operation. 
After a process has invoked an operation, the register may trigger an event that carries the reply from the operation. We say that the process \emph{completes} the operation when this event occurs. 
Moreover, after a process has invoked an operation on a register, the process does not invoke any further operation on that register until the previous operation completes and we say that a correct process accesses the registers in a \emph{sequential} manner.
An operation $o$ \emph{precedes} another operation $o'$ in a sequence of events whenever $o$ completes before $o'$ is invoked. Two operations are \emph{concurrent} if neither one of them precedes the other. 

\begin{definition}[Permissionless SWMR]\label{def:FIXME}
  A protocol for \emph{permissionless single-writer multi-reader} register satisfies the following properties.
  For every league $L$ and every execution tolerated by $L$:
\begin{description}
\item[Termination:] If a correct process $p_i \in L$ invokes an operation on the register, $p_i$ eventually completes the operation.
\item[Validity:] Every $\op{read}$ operation of a correct process in $L$ that is not concurrent with a $\op{write}$ returns the last value written by a correct process in $L$; a $\op{read}$ of a correct process in $L$ concurrent with a $\op{write}$ of a correct process in $L$ may also return the value that is written concurrently.
\end{description}
\end{definition}

In Algorithm~\ref{alg:regular-register-sigs}, the writer $p_w$ waits until receiving \str{ack} messages from all processes in a quorum $Q_w \in \mathscr{Q}(p_w, \mathbb{V})$. 
The reader $p_r$ waits for a \str{value} message with a value/timestamp pair from every process in a quorum $Q_r \in \mathscr{Q}(p_r, \mathbb{V})$.

The function $\op{highestval}(S)$ takes a set of timestamp/value pairs $S$ and returns the value of the pair with the largest timestamp.

Finally, the protocol uses digital signatures, with operations $\op{sign}_i$, invoked by process $p_i$, and $\op{verify}_i$.
In particular, $\op{sign}_i$ takes a message $m \in \{0,1\}^*$ as input and returns a signature $\sigma \in \{0,1\}^*$, while $\op{verify}_i$ takes as input a signature $\sigma$ and a message $m \in \{0,1\}^*$ and returns \true if and only if $p_i$ \op{signed} the message $m$ and obtained $\sigma$, or \false otherwise.

\begin{algo*}[hbt!]
\vbox{
\small
\begin{tabbing}
  xxxx\=xxxx\=xxxx\=xxxx\=xxxx\=xxxx\=MMMMMMMMMMMMMMMMMMM\=\kill
  \textbf{State} \\
  \> \(\var{ts}_w\): sequence number of write operations, stored only by writer~$p_w$ \\
  \> \(\var{id}_r\): identifier of read operations, used only by reader \\
  \> \(\var{ts}, v, \sigma\): current state stored by $p_i$: timestamp, value, signature \\
  \\
  \textbf{upon invocation} \(\op{write}(v)\) \textbf{do}
  \` // if $p_i = p_w$\\
  \> \(\var{ts}_w \gets \var{ts}_w + 1\) \\
  \> \(\sigma \gets \op{sign}_w(\str{write}\|w\|\var{ts}_w\|v)\) \\
  \> send message \([\str{write}, \var{ts}_w, v, \sigma, \mathbb{F}_i]\) through gossip  \\
  \> \textbf{wait for} receiving a gossiped message \([\str{ack}, (P_j, \mathcal{F}_j)]\)\\
  \>\>  from all processes in \( \mathcal{Q}(p_w, \mathbb{V})\) \\
   \> \(\mathbb{V}[j] \gets (P_j, \mathcal{F}_j)\)\\

  \\
  \textbf{upon} receiving a gossiped message \([\str{write}, \var{ts}', v', \sigma',  (P_w, \mathcal{F}_w)]\)  \\
  \>\> from $p_w$ \textbf{do}
  \` // every process \\
  \> \textbf{if} \(\var{ts}' > \var{ts}\) \textbf{then} \\
  \>\> \(\mathbb{V}[w] \gets  (P_w, \mathcal{F}_w)\)\\
  \> \> \((\var{ts}, v, \sigma) \gets (\var{ts}', v', \sigma')\) \\
  \> send message \([\str{ack}, \mathbb{F}[i]]\) through gossip to $p_w$ \\
\\
  \textbf{upon invocation} \(\op{read}\) \textbf{do}
  \` // if $p_i=p_r$ \\
  \> \(\var{id}_r \gets \var{id}_r + 1\) \\
  \> send message \([\str{read}, \var{id}_r , \mathbb{F}[i]]\) through gossip\\
  \> \textbf{wait for} receiving gossiped messages \([\str{value}, r_j, \var{ts}_j, v_j, \sigma_j,  (P_j, \mathcal{F}_j)]\)\\
  \>\>	  from all processes in \(\mathcal{Q}(p_r, \mathbb{V}_r)\) \\
     \>\>\textbf{such that}  \(r_j = \var{id}_r\) \textbf{and}
     \(\op{verify}_w(\sigma_j, \str{write}\|w\|\var{ts}_j\|v_j)\) \\
   \> \(\mathbb{V}[j] \gets  (P_j, \mathcal{F}_j)\)\\
  \> \textbf{return} \( \op{highestval}( \{ (\var{ts}_j, v_j) |~j \in Q_r\} \) \\
  \\
  \textbf{upon} receiving a gossiped message \([\str{read}, r ,  (P_r, \mathcal{F}_r)]\) \\
  \>\> from $p_r$ \textbf{do}
  \` // every process \\
    \> \(\mathbb{V}_i[r] \gets  (P_r, \mathcal{F}_r)\)\\
  \> send message \([\str{value}, r, \var{ts}, v, \sigma, \mathbb{F}_i]\) through gossip to $p_r$
\end{tabbing}
}
\caption{Emulation of a permissionless SWMR regular register (process~$p_i$).}
\label{alg:regular-register-sigs}
\end{algo*}

\begin{theorem}
Algorithm~\ref{alg:regular-register-sigs} implement permissionless SWMR.
\end{theorem}

\begin{proof}
Let us consider a league $L$ and a tolerated execution $e$ with set of faulty processes $A$.
To prove the \emph{termination} property, let us consider a writer $p_w$.
By assumption, process $p_w$ is correct in the league $L$ and, by the availability property of $L$, eventually there exists a quorum $Q_w \in \mathscr{Q}(p_w, \mathbb{F})$ contained in $L\setminus A$.
Therefore, $p_w$ will receive sufficiently many \str{ack} messages and the write will return.
Let $p_r$ be a reader in $L \setminus A$. As above, eventually there exists a quorum $Q_r\in \mathscr{Q}(p_r, \mathbb{F})$ contained in $L \setminus A$.
Because the writer is correct and in the league, all the responses from processes in $Q_r$ satisfy the checks and \op{read} returns.

For the \emph{validity} property, observe that by assumption both the writer $p_w$ and the reader $p_r$ are correct processes in $L\setminus A$.
If the writer $p_w$ \op{writes} to a quorum $Q_w \subseteq \mathscr{Q}(p_w, \mathbb{V})$ for itself, and the reader $p_r$ reads from a quorum $Q_r \subseteq \mathscr{Q}(p_r,\mathbb{V}')$ for itself, with $\mathbb{V}$ and $\mathbb{V}'$ two $A$-resilient views, by the consistency property of $L$ it holds $(Q_w \cap Q_r) \setminus A \neq \emptyset.$
Hence, there is some correct process $p_i \in Q_w \cap Q_r$ that received the most recently written value from $p_w$ and returns it to $p_r$.

Observe that, from the properties of the signature scheme, any value output by \op{read} has been written in some preceding or concurrent \op{write} operation.
\end{proof}

\section{Permissionless Reliable Broadcast}
\label{sec:reliable-broadcast}

In this section we show how the Bracha broadcast~\cite{DBLP:journals/iandc/Bracha87}, protocol that implements Byzantine reliable broadcast, can be adapted to work in our model.
First, we introduce the following definitions and results.

\begin{definition}[Blocking set]
A set $B \subseteq \Pi$ is said to \emph{block} a process $p_i$ if $B$ intersects every slice of $p_i$.
\end{definition}

\begin{definition}[Inductively blocked]
\label{def:inductively-blocking}
  Given a set of processes $B$, the set of processes inductively blocked by $B$, denoted by $B^{+}$, is the smallest set closed under the following rules:
  \begin{enumerate}
    \item $B\subseteq B^{+}$; and
    \item if a process $p_i$ is blocked by $B^{+}$, then $p_i\in B^{+}$.
  \end{enumerate}
\end{definition}

As a consequence of Definition~\ref{def:inductively-blocking}, given an execution, the set $B^{+}$ can be obtained by repeatedly adding to it all the processes that are blocked by $B^{+} \cup B$. Eventually no more processes will be added to $B^{+}$.

Moreover, given an execution $e$ with set of faulty processes $A$, if a league $L$ tolerates $A$, then processes in $L \setminus A$ cannot be inductively blocked by $A$. 
This is shown in the following lemma.

\begin{lemma}
\label{lem:self-sufficient-not-blocked-by-complement}
Let $L$ be a league and $T$ be a set tolerated by $L$.
Then, no process in $L \setminus T$ is inductively blocked by $T$, i.e., $T^{+}\cap (L\setminus T) = \emptyset$.
\end{lemma}

\begin{proof}
Let us assume that $T^{+}\cap (L\setminus T )\neq \emptyset$.
This means that there exists a process $p_i \in L \setminus T$ that is blocked by $T^{+}$, i.e., $T^{+}$ intersects every slice of $p_i$, including the slice contained in the quorum $Q_i \subseteq L\setminus T $ for $p_i$. Clearly, $(L \setminus T) \cap T = \emptyset$, and this means that there exists a set $T'$ with $T' \subseteq T^{+} \setminus T $ such that $T'$ intersects every slice of $p_i$, including the slice contained in the quorum for $p_i$ consisting only of correct processes in $L$. This means that we can find a process $p_j \in T'$ with $p_j \in L \setminus T$ and $p_j$ blocked by $T$.
Since $L$ is a league, process $p_j$ must have a slice in $L \setminus T$.
However, $T$ cannot intersect every slice of $p_j$ because $L \setminus T$ is disjoint from $T$.
We reached a contradiction.
\end{proof}

Intuitively, starting from $A^{+}=\emptyset$, we first consider the processes that are blocked by $A$.
Trivially, every process in $A$ is blocked by $A$, and so $A^{+} = A$.
Moreover, no process in $L\setminus A$ can be blocked by $A$. 
If this was the case, then there would exist a process $p_i \in L\setminus A$ such that $A$ intersected all of its slices, including the slice contained in the quorum $Q_i \subseteq L \setminus A$, which we know to exist due to the availability property of $L$.
So, only processes $p_j$ not in $L \setminus A$ can be blocked by $A$.
Let $p_j$ be such process. This means that $A \cup \{p_j\} \subseteq A^{+}$. 
Now, we can repeat the same reasoning, by considering all the processes blocked by $A \cup \{p_j\}$. Again, no processes in $L\setminus A$ can be blocked by $A \cup \{p_j\}$. 
In fact, if $A \cup \{p_j\}$ blocked a process $p_k \in L\setminus A$, then every slice of $p_k$ would contain $p_j$, including the slice contained in $L\setminus A$.
  However, this would imply that $p_j \in L\setminus A$ which would contradict the fact that $p_j$ is a process not in $L\setminus A$.
  
  In the following theorem we show that if a correct process $p_i$ in a league $L$ is blocked by a set $B$, then $B = B \cup \{p_i\}$ blocks another process $p_j \not\in B \cup A$.
Then, $B' = B \cup \{p_j\}$ blocks another process $p_k \not\in B' \cup A$ and so on, until, eventually, every correct process in the league is blocked.

\begin{theorem}[Cascade theorem]
\label{thm:cascade}
  Consider the quorum function $\mathscr{Q}$, a league $L$, and a set $T \subseteq \Pi$ tolerated by $L$.
  Moreover, let us consider a process $p_i\in L\setminus T$, a $T$-resilient view $\mathbb{V}$ for $p_i$, a quorum $Q_i\in\mathscr{Q}(p_i,\mathbb{V})$, and a set $B \subseteq \Pi$ disjoint from $T$ such that $Q_i\setminus T\subseteq B$.
  Then, either $L\setminus T\subseteq B$ or there exists a process $p_j \not\in B\cup T$ that is blocked by $B$.
\end{theorem}

\begin{proof}
  It suffices to assume by contradiction that $L\setminus (B\cup T)\neq \emptyset$ and that, for every $p_j \not\in B\cup T$, process $p_j$ has a slice disjoint from $B$.
  This implies that $S = \overline{B\cup T}$ is a survivor set of every process $p_j \in S$; since $L\setminus (B\cup T)\neq \emptyset$ , this includes also at least one process $p_j\in L\setminus (B \cup T)$.
  
  Let us consider such a process $p_j \in L \setminus (B \cup T)$ and consider the view $\mathbb{V}'$ for $p_j$ such that: (1) for every $p_k \not\in T$, $\mathbb{V}'[k]=\mathbb{F}[k]$; and (2) for every $p_k\in T$, $\mathbb{V}'[k]=(\emptyset, \{\emptyset\})$. Observe that $\mathbb{V}'$ is a $T$-resilient view for $p_j$. By Lemma~\ref{lem:S-is-quorum}, we have that $S \in \mathscr{Q}(p_j, \mathbb{V}')$ .
  This implies that $S \cap Q_i \subseteq T$. But combined with the fact that $p_j \in L \setminus (B \cup T)$, this contradicts the consistency property of $L$.
\end{proof}

We will see how this theorem has a direct effect on the liveness of permissionless Byzantine reliable broadcast. 


In a Byzantine reliable broadcast, the sender process may \emph{broadcast} a value~$v$ by invoking $\op{r-braodcast}(v)$. The broadcast primitive outputs a value~$v$ through an $\op{r-deliver}(v)$ event.
Moreover, the broadcast primitive presented in this section delivers only one value per instance. 
Every instance has an implicit label and a fixed, well-known sender~$p_s$. 

\begin{definition}[Permissionless Byzantine reliable broadcast]\label{def:r-broadcast}
  A protocol for \emph{permissionless Byzantine reliable broadcast} satisfies the following properties.
  For every league $L$ and every execution tolerated by $L$:
\begin{description}
\item[Validity:] If a correct process $p_s$ \op{r-broadcasts} a value $v$, then all correct processes in $L$ eventually \op{r-deliver} $v$.
\item[Integrity:] For any value $v$, every correct process \op{r-delivers} $v$ at most once.
Moreover, if the sender $p_s$ is correct and the receiver is correct and in $L$, then $v$ was previously \op{r-broadcast} by $p_s$.
\item[Consistency:] If a correct process in $L$ \op{r-delivers} some value~$v$ and another correct process in $L$ \op{r-delivers} some value~$v'$, then $v=v'$.
\item[Totality:]  If a correct process in $L$ \op{r-delivers} some value $v$, then all correct processes in $L$ eventually \op{r-deliver} some value.
\end{description}
\end{definition}

We implement this primitive in Algorithm~\ref{alg:reliable-broadcast}, which is derived from Bracha broadcast~\cite{DBLP:journals/iandc/Bracha87} but differs in some aspects. 

In principle, the protocol follows the original one, but does not use one global quorum system known to all processes.  Instead, the correct processes implicitly use the same quorum function $\mathscr{Q}$ (Definition~\ref{def:quorum-function}), of which they initially only know their own entry in $\mathscr{Q}$.  They discover the quorums of other processes during the execution.

Because of the permissionless nature of our model, we consider a best-effort gossip primitive to disseminate messages among processes instead of point-to-point messages.

A crucial element of Bracha's protocol is the ``amplification'' step, when a process receives $f+1$ \str{ready} messages with some value~$v$, with $f$ the number of faulty processes in an execution, but has not sent a \str{ready} message yet.  Then it also sends a \str{ready} message with~ $v$.  This generalizes to receiving the same \str{ready} message with value $v$ from a \emph{blocking set} for $p_i$ and is crucial for the \emph{totality} property.

Finally, we introduce the \str{any} message as a message sent by a process $p_i$ that is blocked by two sets carrying two different values $v$ and~$v'$.
The reason for this new message lies in the consistency property of $L$: given an execution $e$ with set of faulty processes $A$ tolerated by $L$, the consistency property of $L$ implies that any two quorums of any two correct processes in $L$ have some correct process in common. 
Quorum intersection is then guaranteed only for correct processes in $L$ and nothing is assured for correct processes outside $L$, which might gossip different values received by non-intersecting quorums. 
In particular, if a correct process $p_i$ is blocked by a set containing a value $v$ and later is blocked by a set containing a value $v' \neq v$, then $p_i$ gossips an \str{any} message containing~$*$.
\str{any} messages are then ignored by correct processes in~$L$.
As we show in the Theorem~\ref{thm:rb}, correct process in $L$ cannot be blocked by sets containing different values.

\begin{algo*}[hbt!]
\vbox{
\small
\begin{tabbing}
  xxxx\=xxxx\=xxxx\=xxxx\=xxxx\=xxxx\=MMMMMMMMMMMMMMMMMMM\=\kill
  \textbf{State} \\
  \> \(\var{sent-echo} \gets \false\): indicates whether $p_i$ has gossiped $\str{echo}$ \\
  \> \(\var{echos}[j] \gets [\perp]\): collects the received $\str{echo}$ messages from other processes \\
  \> \(\var{sent-ready} \gets \false\): indicates whether $p_i$ has gossiped $\str{ready}$ \\
  \> \(\var{readys}[j]  \gets [\perp]\): collects the received $\str{ready}$ messages from other processes \\
      \> \(\var{sent-any} \gets \false\): indicates weather $p_i$ has gossiped [\str{any}, $*$, $\mathbb{F}[i]$]\\
  \> \(\var{delivered} \gets \false\): indicates whether $p_i$ has delivered a value\\
  \> \(\mathbb{V}[j] \gets \text{ if } i=j\text{ then }\mathbb{F}[i]\text{ else }\bot\): the current view of $p_i$\\
  \\
  \textbf{upon invocation} \(\op{r-broadcast}(v)\) \textbf{do} \\
  \> send message [\str{send}, $v$, $\mathbb{F}[s]$]  through gossip  \` // only sender $p_s$ \label{}\\
  \\
  \textbf{upon} receiving a gossiped message  [\str{send}, $v$, $(P_s, \mathcal{F}_s)$] from $p_s$
    \textbf{and} \(\neg \var{sent-echo}\) \textbf{do} \\
  \> \(\var{sent-echo} \gets \true\) \\
  \>  \(\mathbb{V}[s] \gets (P_s, \mathcal{F}_s)\)\\
  \> send message  [\str{echo}, $v$, $\mathbb{F}[i]$] through gossip \\
  \\
  \textbf{upon} receiving a gossiped message  [\str{echo}, $v$, $(P_j, \mathcal{F}_j)$ from \(p_j\) \textbf{do} \\
  \> \textbf{if} \(\var{echos}[j] = \perp \) \textbf{then} \\
    \>\>  \(\mathbb{V}[j] \gets (P_j, \mathcal{F}_j)\)\\
  \> \> \(\var{echos}[j] \gets v\) \\
  \\
  \textbf{upon exists} \(v \not= \perp \) \textbf{such that}
     \( \{p_j \in \Pi |~\var{echos}[j] = v\} \in \mathscr{Q}(p_i,\mathbb{V}) \)
     \textbf{and} \(\neg \var{sent-ready}\) \textbf{do}\\
  \> \(\var{sent-ready} \gets \true\) \\
  \> send message  [\str{ready}, $v$, $\mathbb{F}[i]$] through gossip\\
  \\
    \textbf{upon} receiving a gossiped message [\str{ready}, $v$, $(P_j, \mathcal{F}_j)$] from \(p_j\)
     \textbf{do} \\
       \> \textbf{if} \(\var{readys}[j] = \perp\) \textbf{then} \\
       \>\>  \(\mathbb{V}[j] \gets (P_j, \mathcal{F}_j)\)\\
  \> \> \(\var{readys}[j] \gets v\) \\
  \\
     \textbf{upon exists} \(v \not= \perp \) \textbf{such that}
     \( \{p_j \in \Pi |~\var{readys}[j] = v\} \) blocks $p_i$
     \textbf{and} \(\neg \var{sent-ready}\) \textbf{do} \\
  \> \(\var{sent-ready} \gets \true\) \\
  \> send message [\str{ready}, $v$, $\mathbb{F}[i]$] through gossip \\
  \\
       \textbf{upon exists} \(v' \not= \perp\) \textbf{such that}
     \( \{p_j \in \Pi |~\var{readys}[j] = v'\} \) blocks $p_i$
     \textbf{and} \(\var{readys}[i] = v \) \textbf{and} \\
     \>\> $v \neq v'$ \textbf{and}  \(\var{sent-ready}\) \textbf{and} \(\neg\var{sent-any}\) \textbf{do} \\
  \> \(\var{sent-any} \gets \true\) \\
  \> send message [\str{any}, $*$, $\mathbb{F}[i]$] through gossip \\
  \\
      \textbf{upon} receiving a gossiped message [\str{any}, $*$, $(P_j, \mathcal{F}_j)$] from \(p_j\)
     \textbf{do} \\
        \>  \(\mathbb{V}[j] \gets (P_j, \mathcal{F}_j)\)\\
  \> \(\var{readys}[j] \gets *\) \\
  \\
  \textbf{upon exists} \(v \not= \perp \) \textbf{such that}
     \( \{ p_j \in \Pi |~\var{readys}[j]  = v\} \in \mathscr{Q}(p_i,\mathbb{V}) \)
     \textbf{and} \(\neg \var{delivered}\) \textbf{do} \\
  \> \(\var{delivered} \gets \true\) \\
  \> \textbf{output} \(\op{r-deliver(v)}\)

\end{tabbing}
}
\caption{Permissionless Byzantine reliable broadcast protocol for process~$p_i$, with sender~$p_s$}
\label{alg:reliable-broadcast}
\end{algo*}

\begin{theorem}
\label{thm:rb}
Algorithm~\ref{alg:reliable-broadcast} implements permissionless Byzantine reliable broadcast.
\end{theorem}

\begin{proof}
Observe that all the properties assume the existence of a league $L$ and an execution $e$ with set of faulty processes $A$ tolerated by $L$.

Let us start with the \emph{validity} property.
Since the sender $p_s$ is correct and from the availability property of $L$, every correct process $p_i$ in $L$ eventually receives a quorum $Q_i$ for itself of \str{echo} messages containing the value $v$ sent from $p_s$ and updates its view $\mathbb{V}$ according to the views received from every process in $Q_i$.

Then, $p_i$ gossips [\str{ready}, $v$, $\mathbb{F}[i]$] containing the value $v$ and its current view $\mathbb{F}[i]$ unless $\var{sent-ready}= \str{true}$. If $\var{sent-ready}= \str{true}$ then $p_i$ already gossiped [\str{ready}, $v$, $\mathbb{F}[i]$]. 

Observe that there exists a unique value $v$ such that if a correct process in $L$ sends a \str{ready} message, 	this message contains $v$. 
In fact, if a process $p_i \in L\setminus A$ sends a \str{ready} message, either it does so after receiving a quorum $Q_i$ for itself of \str{echo} messages containing $v$ or after being blocked by a set of processes that received \str{ready} messages containing $v$.

In the first case, if a correct process $p_i$ in $L$ receives a quorum $Q_i$ for itself of \str{echo} messages containing $v$ and another correct process $p_j$ in $L$ receives a quorum $Q_j$ for itself of \str{echo} messages containing $v'$, by the consistency property of $L$, $v=v'$ and both send a \str{ready} message containing the same $v$.

In the second case, first observe that by Lemma~\ref{lem:self-sufficient-not-blocked-by-complement} we know that $p_i \in L\setminus A$ cannot be inductively blocked by processes in $A$.
Moreover, correct processes in $L$ cannot be blocked by sets containing different values.
If this was the case, then there would exist two correct processes $p_i$ and $p_j$ in $L$ and two slices of $p_i$ and $p_j$, respectively, in $L\setminus A$ containing two correct processes in $L$ that received two different values $v$ after \str{echo}.
Again, by the consistency property of $L$, this is not possible. Hence, every correct process $p_j$ in $L$ gossips
[\str{ready}, $v$, $\mathbb{F}[j]$].
Eventually, every correct process $p_i$ in $L$ receives a quorum for itself containing [\str{ready}, $v$, $(P_j,\mathcal{F}_j)$] messages and \op{r-delivers}~$v$.

The first part of the \emph{integrity} property is ensured by the $\var{delivered}$ flag. 
For the second part observe that, by assumption, the receiver $p_i$ is correct and in $L$. This implies that the quorum for $p_i$ used to reach a decision contains some correct processes that have gossiped \str{echo} containing a value $v$ they received from $p_s$. 

For the \emph{totality} property, let us assume that a correct process $p_i \in L$ \op{r-delivered} some value $v$.
If $p_i \in L\setminus A$ \op{r-delivered} some value $v$, then it has received \str{ready} messages containing $v$ from a quorum $Q_i$ for itself.
From Theorem~\ref{thm:cascade} we know that exists a set $B$ such that $Q_i \setminus A \subseteq B$ and either $L \setminus A \subseteq B$ or $B$ blocks at least a process $p_j \in L \setminus (B \cup A)$ in an $A$-resilient view $\mathbb{V}'$ for $p_j$. In the latter case, $p_j$ gossips a \str{ready} message containing $v$ and $B$ becomes $B \cup \{p_j\}$.
Observe that, by assumption, if a correct process receives a gossiped message, then eventually every other correct process receives it too.
Eventually, $L\setminus A$ is covered by $B$ and this means that every correct process in $L$ is blocked with the same value~$v$. 

Moreover, observe that given two correct processes not in $L$, they may become ready for different values received from non-intersecting quorums of \str{echo} messages. Because of this, if a correct process $p_j \not\in L$ observes a blocking set $B$ containing a value $v'$ different from a value $v$ that has previously gossiped in a \str{ready} message and such that $\var{sent-any}=\false$, process $p_j$ gossips an \str{any} message containing the value $*$. Eventually every correct process $p_i$ in $L$ receives a quorum $Q_i$ for itself of [\str{ready}, $v$, $(P_j,\mathcal{F}_j)$] messages and it \op{r-delivers} $v$.

Finally, for the \emph{consistency} property notice that by the consistency property of $L$, every two quorums $Q_i$ and $Q_j$ of any two correct processes $p_i$ and $p_j$ in $L$ intersect in some correct process $p_k$.
Process $p_k$ could then be outside $L$.
If $p_k \not\in L$ then, as seen for the totality property, it can be blocked by sets containing different values.
If this is the case then $p_k$ gossips an \str{any} message.
Correct processes in $L$ then ignore the values received from $p_k$ and wait until receiving a quorum unanimously containing the same value $v$.
Observe that, because $L$ tolerates $A$, by availability property of $L$ every correct process in $L$ eventually receives a quorum made by correct processes in $L$.
The consistency property then follows.
\end{proof}

\section{Related Work}
\label{sec:related-work}

A \emph{fail-prone system}~\cite{DBLP:journals/dc/MalkhiR98}, also called adversary structure~\cite{DBLP:journals/joc/HirtM00}, is a well-adopted way to describe the failure assumptions in a distributed system.
This is a collection of subsets of participants in the system that may fail together, and that are tolerated to fail, in a given execution.
Fail-prone systems implicitly define \emph{Byzantine quorum systems}~\cite{DBLP:journals/dc/MalkhiR98} which are used to ensure consistency and availability to distributed fault-tolerant protocols in the presence of arbitrary failures.

Originally, fail-prone systems have been expressed globally, shared by every participant in the system.
Damg{\aa}rd \emph{et al.}~\cite{DBLP:conf/asiacrypt/DamgardDFN07} introduce the notion of \emph{asymmetric} fail-prone system in which every participant in the system subjectively selects its own fail-prone system, allowing for a more flexible model and where the guarantees of the system are derived from personal assumptions. This is the \emph{asymmetric-trust model}~\cite{DBLP:conf/opodis/CachinT19}.
Processes in this model are classified in three different types, \emph{faulty}, \emph{naive}, and \emph{wise} and this characterization is done with respect to an execution.
In particular, given an execution $e$ with faulty set $F$, a process $p_i$ is faulty if it belongs to $F$, $p_i$ is naive if it does not have $F$ in its subjective fail-prone system and $p_i$ is wise if $F$ is contained in its subjective fail-prone system.
Properties of protocols are then guaranteed for wise processes and, in some cases~\cite{DBLP:conf/opodis/CachinT19, DBLP:conf/esorics/CachinZ21}, for a subset of the wise processes called \emph{guild}.

Cachin and Tackmann~\cite{DBLP:conf/opodis/CachinT19} introduce \emph{asymmetric Byzantine quorum systems}, a generalization of the original Byzantine quorum in the asymmetric-trust model.
They show how to implement register abstraction and broadcast primitives using asymmetric Byzantine quorum systems.
An asynchronous consensus protocol has subsequently been devised by Cachin and Zanolini~\cite{DBLP:conf/esorics/CachinZ21}. Moreover, they extended the knowledge about the guild and about the relation between naive and wise processes in protocols with asymmetric trust.

As a basis for the \emph{Stellar consensus protocol}, Mazi{\`e}res~\cite{Mazieres2015TheSC} introduces a new model called \emph{federated Byzantine agreement}~(FBA) in which participants may also lie.
Here, every participant declares \emph{quorum slices} -- a collection of trusted sets of processes sufficient to convince the particular participant of agreement.
These slices make a \emph{quorum}, a set of participants that contains one slice for each member and sufficient to reach agreement.
All quorums constitute a \emph{federated Byzantine quorum system}~(FBQS).
In this model, even if the processes do not a priori choose intersecting quorums as in the classic model~\cite{DBLP:journals/dc/MalkhiR98} or as in the asymmetric one~\cite{DBLP:conf/asiacrypt/DamgardDFN07, DBLP:conf/opodis/CachinT19}, an intersection property among quorums is later required for the analysis of the Stellar consensus protocol.

Garc{\'{\i}}a{-}P{\'{e}}rez and Gotsman~\cite{DBLP:conf/opodis/Garcia-PerezG18} study the theoretical foundations of a FBQS, build a link between FBQS and the classical Byzantine quorum systems and show the correctness of broadcast abstractions over federated quorum systems. Moreover, they investigate decentralized quorum constructions by means of FBQS.
Finally, they propose the notion of subjective dissemination quorum system, where different participants may have different Byzantine quorum systems and where there is a system-wide intersection property.
FBQS are a way towards an extension of quorum systems in a permissionless setting.

Losa \emph{et al.}~\cite{DBLP:conf/wdag/LosaGM19} introduce \emph{personal Byzantine quorum systems} (PBQS) by removing from FBQS the requirement of a system-wide intersection among quorums.
This might lead to disjoint \emph{consensus clusters} in which safety and liveness are guaranteed in each of them, separately, in a given execution.
Moreover, they abstract the Stellar Network as an instance of PBQS and use a PBQS to solve consensus.

\section{Conclusions}
\label{sec:conclusions}

This work introduces a new way of specifying trust assumptions among processes in a permissionless setting: processes not only make assumptions about failures, but also make assumptions about the assumptions of other processes.
This leads to formally define the notions of {permissionless fail-prone system} and  permissionless quorum system and to design protocols to solve known synchronization problems such as Byzantine reliable broadcast.

We introduce the notion of league, a set of processes for which consistency and availability properties hold for a given quorum function. Properties of our protocols are guaranteed assuming the existence of a league.

As a future work we plan to generalize known consensus protocols such as, for example, PBFT~\cite{DBLP:journals/tocs/CastroL02}, to work in our permissionless model. We believe that, by assuming the existence of a league $L$, properties of consensus protocols can be guaranteed to every correct process~in~$L$.

\section*{Acknowledgments}

The authors thank anonymous reviewers for helpful feedback.

This work has been funded by the Swiss National Science Foundation (SNSF)
under grant agreement Nr\@.~200021\_188443 (Advanced Consensus Protocols).

\bibliography{references, dblpbibtex}

\begin{thebibliography}{10}

\bibitem{DBLP:conf/opodis/Amores-SesarCM20}
Ignacio Amores{-}Sesar, Christian Cachin, and Jovana Micic.
\newblock Security analysis of ripple consensus.
\newblock In {\em {OPODIS}}, volume 184 of {\em LIPIcs}, pages 10:1--10:16.
  Schloss Dagstuhl - Leibniz-Zentrum f{\"{u}}r Informatik, 2020.

\bibitem{DBLP:journals/iandc/Bracha87}
Gabriel Bracha.
\newblock Asynchronous byzantine agreement protocols.
\newblock {\em Inf. Comput.}, 75(2):130--143, 1987.

\bibitem{DBLP:books/daglib/0025983}
Christian Cachin, Rachid Guerraoui, and Lu{\'{\i}}s E.~T. Rodrigues.
\newblock {\em Introduction to Reliable and Secure Distributed Programming
  {(2.} ed.)}.
\newblock Springer, 2011.

\bibitem{DBLP:conf/opodis/CachinT19}
Christian Cachin and Bj{\"{o}}rn Tackmann.
\newblock Asymmetric distributed trust.
\newblock In {\em {OPODIS}}, volume 153 of {\em LIPIcs}, pages 7:1--7:16.
  Schloss Dagstuhl - Leibniz-Zentrum f{\"{u}}r Informatik, 2019.

\bibitem{DBLP:conf/esorics/CachinZ21}
Christian Cachin and Luca Zanolini.
\newblock Asymmetric asynchronous byzantine consensus.
\newblock In {\em DPM/CBT@ESORICS}, volume 13140 of {\em Lecture Notes in
  Computer Science}, pages 192--207. Springer, 2021.

\bibitem{DBLP:journals/tocs/CastroL02}
Miguel Castro and Barbara Liskov.
\newblock Practical byzantine fault tolerance and proactive recovery.
\newblock {\em {ACM} Trans. Comput. Syst.}, 20(4):398--461, 2002.

\bibitem{DBLP:conf/asiacrypt/DamgardDFN07}
Ivan Damg{\aa}rd, Yvo Desmedt, Matthias Fitzi, and Jesper~Buus Nielsen.
\newblock Secure protocols with asymmetric trust.
\newblock In {\em {ASIACRYPT}}, volume 4833 of {\em Lecture Notes in Computer
  Science}, pages 357--375. Springer, 2007.

\bibitem{DBLP:conf/opodis/Garcia-PerezG18}
{\'{A}}lvaro Garc{\'{\i}}a{-}P{\'{e}}rez and Alexey Gotsman.
\newblock Federated byzantine quorum systems.
\newblock In {\em {OPODIS}}, volume 125 of {\em LIPIcs}, pages 17:1--17:16.
  Schloss Dagstuhl - Leibniz-Zentrum f{\"{u}}r Informatik, 2018.

\bibitem{DBLP:journals/joc/HirtM00}
Martin Hirt and Ueli~M. Maurer.
\newblock Player simulation and general adversary structures in perfect
  multiparty computation.
\newblock {\em J. Cryptol.}, 13(1):31--60, 2000.

\bibitem{DBLP:conf/icdcs/JunqueiraM03}
Flavio~Paiva Junqueira and Keith Marzullo.
\newblock Synchronous consensus for dependent process failure.
\newblock In {\em {ICDCS}}, pages 274--283. {IEEE} Computer Society, 2003.

\bibitem{DBLP:conf/wdag/LosaGM19}
Giuliano Losa, Eli Gafni, and David Mazi{\`{e}}res.
\newblock Stellar consensus by instantiation.
\newblock In {\em {DISC}}, volume 146 of {\em LIPIcs}, pages 27:1--27:15.
  Schloss Dagstuhl - Leibniz-Zentrum f{\"{u}}r Informatik, 2019.

\bibitem{DBLP:conf/ccs/MalkhiN019}
Dahlia Malkhi, Kartik Nayak, and Ling Ren.
\newblock Flexible byzantine fault tolerance.
\newblock In {\em {CCS}}, pages 1041--1053. {ACM}, 2019.

\bibitem{DBLP:journals/dc/MalkhiR98}
Dahlia Malkhi and Michael~K. Reiter.
\newblock Byzantine quorum systems.
\newblock {\em Distributed Comput.}, 11(4):203--213, 1998.

\bibitem{DBLP:journals/siamcomp/MalkhiRW00}
Dahlia Malkhi, Michael~K. Reiter, and Avishai Wool.
\newblock The load and availability of byzantine quorum systems.
\newblock {\em {SIAM} J. Comput.}, 29(6):1889--1906, 2000.

\bibitem{Mazieres2015TheSC}
David Mazi{\`e}res.
\newblock The {Stellar} consensus protocol: A federated model for
  {Internet}-level consensus.
\newblock Stellar, available online,
  \url{https://www.stellar.org/papers/stellar-consensus-protocol.pdf}, 2016.

\bibitem{DBLP:conf/wdag/PassS17}
Rafael Pass and Elaine Shi.
\newblock Hybrid consensus: Efficient consensus in the permissionless model.
\newblock In {\em {DISC}}, volume~91 of {\em LIPIcs}, pages 39:1--39:16.
  Schloss Dagstuhl - Leibniz-Zentrum f{\"{u}}r Informatik, 2017.

\bibitem{DBLP:journals/corr/SheffRM14}
Isaac~C. Sheff, Robbert van Renesse, and Andrew~C. Myers.
\newblock Distributed protocols and heterogeneous trust: Technical report.
\newblock {\em CoRR}, abs/1412.3136, 2014.

\bibitem{DBLP:conf/opodis/SheffWRM20}
Isaac~C. Sheff, Xinwen Wang, Robbert van Renesse, and Andrew~C. Myers.
\newblock Heterogeneous paxos.
\newblock In {\em {OPODIS}}, volume 184 of {\em LIPIcs}, pages 5:1--5:17.
  Schloss Dagstuhl - Leibniz-Zentrum f{\"{u}}r Informatik, 2020.

\bibitem{DBLP:journals/dc/SrikanthT87}
T.~K. Srikanth and Sam Toueg.
\newblock Simulating authenticated broadcasts to derive simple fault-tolerant
  algorithms.
\newblock {\em Distributed Comput.}, 2(2):80--94, 1987.

\end{thebibliography}

\end{document}